\documentclass[a4paper,11pt]{article}

\usepackage{jheppub}
\usepackage[T1]{fontenc}
\usepackage{hyperref}
\usepackage{graphicx}
\usepackage{amsmath}
\usepackage{epsfig}
\usepackage{subfig}
\usepackage{xcolor}
\usepackage{natbib}
\usepackage{multirow}

\def\pslash{p\!\!\!\slash}
\def\pbarslash{\bar{p}\!\!\!\slash}

\def\Pslash{P\!\!\!\slash}

\def\OMIT#1{}

\newcommand{\beq}{\begin{equation}}
\newcommand{\eeq}{\end{equation}}
\newcommand{\bqa}{\begin{eqnarray}}
\newcommand{\eqa}{\end{eqnarray}}

\newcommand{\bseq}{\begin{subequations}}
\newcommand{\eseq}{\end{subequations}}

\title{The next-to-next-to-leading-order QCD corrections to $e^+e^-\to \eta_c/\chi_{cJ}+\gamma$ at B factories}
\author[a]{Cong Li,}
\author[a]{Wen-Long Sang,}
\author[b]{Hong-Fei Zhang}

\affiliation[a]{School of Physical Science and Technology, Southwest University, Chongqing 400700, China}
\affiliation[b]{College of Big Data Statistics, Guizhou University of Finance and Economics, Guiyang, 550025, China}

\emailAdd{lc312321@163.com}
\emailAdd{wlsang@swu.edu.cn}
\emailAdd{hfzhang@mail.gufe.edu.cn}

\abstract{
We investigate the processes $e^+e^-\to \eta_c+\gamma$ and $e^+e^-\to \chi_{cJ}+\gamma$ at B factories  within the NRQCD factorization framework, computing the corresponding helicity amplitudes through $\mathcal{O}(\alpha_s^2)$.
The short-distance coefficients are obtained as series expansions in $r=\frac{4m_c^2}{s}$ around $r=0, 1/3, 2/3, 1$, using the method of differential equations. 
By combining the expansions from all four points, we construct composite asymptotic expressions that reproduce the exact results accurately over the full range $0 \leq r\leq 1$, with relative errors below $0.1\%$ over most of the domain and remaining under $1\%$ elsewhere.
Analytic expressions for the leading and next-to-leading logarithmic terms are extracted in the limit $r\to 0$. Using these results, we compute the unpolarized cross sections and observe that the perturbative corrections are small for $\chi_{c0}+\gamma$, moderate for $\chi_{c1}+\gamma$, and substantial for $\eta_c+\gamma$ and $\chi_{c2}+\gamma$. Theoretical prediction for $\chi_{c1}+\gamma$ is consistent with the {\tt Belle} measurement within $2\sigma$, showing good agreement between theory and experiment.
We also predict the angular distribution parameters $\alpha^H_\theta$, which are insensitive to NRQCD matrix elements and exhibit small theoretical uncertainties. These parameters further display good stability across different perturbative orders. With the high luminosity anticipated at {\tt Belle 2}, future experimental measurements will thus provide a clear test of NRQCD factorization. }


\begin{document}

\maketitle

\bibliographystyle{JHEP}

\section{Introduction}\label{intro}

Due to their high luminosity and clean experimental environment, electron-positron colliders, particularly B factories, provide an ideal platform for studying exclusive charmonium production. Several measurements of such processes have been reported~\cite{Abe:2002rb,Aubert:2005tj,Pakhlov:2009nj,Belle:2018jqa} at B factories. Searches for $e^+e^-\to \chi_{c1,2}+\gamma$ and $e^+e^-\to \eta_c+\gamma$ have also been conducted at BESIII~\cite{BESIII:2014uzr,BESIII:2017nty}.

Among various charmonium production channels, the exclusive processes $e^+ e^-\to\eta_c+\gamma$ and $e^+e^-\to \chi_{cJ}+\gamma$ (with $J=0,1,2$) are of particular interest. Their relatively simple structure makes them excellent probes for testing non-relativistic QCD (NRQCD) factorization~\cite{Bodwin:1994jh}. Theoretical studies of these processes within the NRQCD framework have been carried out extensively.

The leading-order (LO) cross sections for $e^+ e^-\to\eta_c/\chi_{cJ}+\gamma$ at B factories were first computed in Ref.~\cite{Chung:2008km}. Subsequent work incorporated next-to-leading-order (NLO) QCD corrections~\cite{Sang:2009jc,Li:2009ki}, and the asymptotic behaviors of the NLO results were later reproduced using light-cone factorization in Ref.~\cite{Wang:2013ywc}. Resummation of large logarithms $\ln\frac{s}{m_c^2}$ at next-to-leading-logarithmic accuracy was further performed in Ref.~\cite{Chung:2019ota}.
Relativistic corrections were investigated for $\eta_c+\gamma$ in Ref.~\cite{Sang:2009jc,Fan:2012dy,Xu:2014zra} and for $\chi_{cJ}+\gamma$ in Refs.~\cite{Xu:2014zra,Brambilla:2017kgw}.

The next-to-next-to-leading-order (NNLO) QCD correction to $e^+ e^-\to\eta_c+\gamma$ was analytically computed by Chen et al. in Ref.~\cite{Chen:2017pyi}, with results expressed in terms of Goncharov polylogarithms, Chen's iterated integrals, and elliptic functions. This result was later revisited numerically in Ref.~\cite{Yu:2020tri}. The NNLO corrections for $e^+ e^-\to\chi_{cJ}+\gamma$ were presented in Ref.~\cite{Sang:2020fql}.

It is also noteworthy that the electromagnetic form factor for $\gamma^*\to \eta_b+\gamma$ was analyzed in Ref.~\cite{Jia:2008ep} by bridging NRQCD and light-cone factorization, where leading logarithms of $\ln\frac{s}{m_c^2}$ were resummed to all orders in $\alpha_s$. Their formalism can be adapted to the case of $e^+e^-\to \eta_c+\gamma$. Related studies based on light-cone approach can also be found in the literature~\cite{Braguta:2010mf}.
Additionally, cross sections for these processes have been evaluated at $Z$-factory energies in Refs.~\cite{Trunin:2024ibx,Liao:2021ifc,Chen:2013itc,Chen:2013mjb}, as well as at BESIII energies in Refs.~\cite{Chao:2013cca, Li:2013nna}.
In related work, NNLO QCD corrections to the electromagnetic form factors of $\eta_c$ and $\eta_b$ have been studied in Refs.~\cite{Feng:2015uha,Wang:2018lry,Babiarz:2025agk}.

In this work, we compute the helicity amplitudes for $e^+e^-\to \eta_c+\gamma$ and $e^+e^-\to \chi_{cJ}+\gamma$ up to $\mathcal{O}(\alpha_s^2)$. Our goal is to derive asymptotic expansions for these amplitudes in the variable $r\equiv \frac{4m_c^2}{s}$ around four expansion points, ensuring that the resulting series accurately approximate the exact results across the full physical range $0\leq r \leq 1$. These expressions allow us to predict not only the unpolarized cross sections but also the azimuthal angular distributions of the produced charmonia, thereby offering more detailed observables for experimental tests of NRQCD factorization. Furthermore, we extract analytic expressions of the large logarithmic terms from the asymptotic expansion around $r=0$, which can support further theoretical investigations.     

The paper is organized as follows. In Sec.~\ref{sec:theo:fram}, we introduce the theoretical framework for computing the helicity amplitudes and cross sections. The computational techniques employed are outlined in Sec.~\ref{sec:com:tec}. In Sec.~\ref{sec:asy:exp}, we present analytic expressions for several lower-order coefficients in the expansion around $r=0$. The convergence of the asymptotic expansions is analyzed in Sec.~\ref{sec:conv} by comparing them with exact results. Phenomenological predictions and related discussions are presented in Sec.~\ref{sec:phe}. Finally, a summary is given in Sec.~\ref{sec:sum}. Explicit expressions for the asymptotic expansion at $r=0$ up to $\mathcal{O}(r^1)$ are provided in Appendices~\ref{app:1loop} and \ref{app:2loop}.

\section{Theoretical framework~\label{sec:theo:fram}}
\subsection{Helicity amplitudes and cross sections}

It is convenient to express the differential cross sections for $e^+e^-\to H+\gamma$ (with $H=\eta_c, \chi_{cJ}$)  in terms of the differential decay width of a timelike photon:
\begin{eqnarray}
\label{pamp}
\frac{d\sigma[e^+ e^- \to H(\lambda_1) + \gamma(\lambda_2)]}{d \cos\theta}&=&\frac{2\pi\alpha}{s^{\frac{3}{2}}}\Sigma_a \frac{d\Gamma[\gamma^*(S_Z)\to H(\lambda_1)+\gamma(\lambda_2)]}{d \cos\theta} 
\nonumber\\
&=&\frac{2\pi\alpha}{s^{\frac{3}{2}}}\frac{|\bf P_1|}{16\pi s }|d^{1}_{S_{z},\lambda}(\theta)|^2|\mathcal{A}^{H}_{\lambda_1 , \lambda_2}|^2 
\nonumber\\
&=&\frac{\alpha}{8s^2}\left(\frac{|\bf P_1|}{\sqrt{s} }\right)|\mathcal{A}^{H}_{\lambda_1 , \lambda_2}|^2 \times \big\{^{\frac{1+\cos^2 \theta}{2}, ~~~     \lambda=\pm 1}_{1-\cos^2 \theta,  ~~~~   \lambda=0},
\end{eqnarray}
where $\theta$ is the polar angle between the outgoing $H$ and the electron beam direction, $\lambda_1$ and $\lambda_2$ are the helicities of $H$ and $\gamma$, respectively, and $S_z$ is the magnetic quantum number of the timelike photon. The quantity $\mathcal{A}^{H}_{\lambda_1 , \lambda_2}$ denotes the corresponding helicity amplitude. The angular distribution is completely governed by the Wigner 
$d$-function $d_{S_z,\lambda}^1(\theta)$, with $\lambda \equiv \lambda_1 - \lambda_2$. Note that angular‑momentum conservation restricts $|\lambda| \leq 1$. The magnitude of the three-momentum of $H$ is given by 
\begin{eqnarray}
|{\bf P_{1}}|=\frac{\lambda^{\frac{1}{2}}(s,M_{H}^2, 0)}{2\sqrt{s}}, \quad \lambda(x,y, z)=x^2+y^2+z^2-2(xy+xz+yz).
\end{eqnarray}

From parity invariance, we obtain:
\begin{eqnarray}
\label{parityamp}
\mathcal{A}^{\eta_c}_{\lambda_1 , \lambda_2}=- \mathcal{A}^{\eta_c}_{-\lambda_1 , -\lambda_2},\quad \mathcal{A}^{\chi_{cJ}}_{\lambda_1 , \lambda_2}=(-1)^J \mathcal{A}^{\chi_{cJ}}_{-\lambda_1 , -\lambda_2}.
\end{eqnarray}
As implied by Eq. (\ref{parityamp}), the number of independent helicity amplitudes is $1, 1, 2, 3$ for the $\eta_c+\gamma$, $\chi_{c0}+\gamma$, $\chi_{c1}+\gamma$, and $\chi_{c2}+\gamma$ channels, respectively. Moreover, the helicity amplitudes exhibit the asymptotic scaling
\begin{equation}
\mathcal{A}^H_{\lambda_1, \lambda_2}\propto r^{\tfrac12(1+ |\lambda_1|)}
\end{equation}
in the limit $s\to \infty$, where $r= 4m_c^2/s$.
The factor $r^{1/2}$ reflects the large momentum transfer required for a heavy‑quark pair to form a quarkonium state with small relative momentum, while the remaining powers follow from the helicity‑selection rule.

Summing over all possible helicity states of $H$ and $\gamma$ in Eq.~\eqref{pamp} yields the unpolarized differential cross section:
\begin{eqnarray}
\label{npamp}
\frac{d\sigma[e^+ e^- \to  H+\gamma]}{d \cos\theta}=\frac{\alpha}{8s^2}\left(\frac{|\mathbf{P_1} |}{\sqrt{s}}\right) A^{H}(1+\alpha^{H}_{\theta} \cos^2\theta),
\end{eqnarray}
where $\alpha^{H}_{\theta}$ is the angular-distribution parameter that characterizes the profile of the angular distribution of $H$ and satisfies $|\alpha^{H}_{\theta}| \le 1$.

The explicit expressions for $A^H$ and $\alpha^H_\theta$ read
\begin{eqnarray}
\label{eq:ang:dis}
A^{\eta_c}&=&|\mathcal{A}^{\eta_c}_{0,1}|^2,~~~~\alpha^{\eta_c}_\theta=1,\nonumber\\
A^{\chi_{c0}}&=&|\mathcal{A}^{\chi_{c0}}_{0,1}|^2,~~~~\alpha^{\chi_{c0}}_\theta=1,\nonumber\\
A^{\chi_{c1}}&=&2|\mathcal{A}^{\chi_{c1}}_{1 , 1}|^2+|\mathcal{A}^{\chi_{c1}}_{0,1}|^2,~~~~\alpha^{\chi_{c1}}_\theta=\frac{|\mathcal{A}^{\chi_{c1}}_{0,1}|^2-2|\mathcal{A}^{\chi_{c1}}_{1 , 1}|^2}{|\mathcal{A}^{\chi_{c1}}_{0,1}|^2+2|\mathcal{A}^{\chi_{c1}}_{1 , 1}|^2},\nonumber\\
A^{\chi_{c2}}&=&|\mathcal{A}^{\chi_{c2}}_{2,1}|^2+2|\mathcal{A}^{\chi_{c2}}_{1 , 1}|^2+|\mathcal{A}^{\chi_{c2}}_{0,1}|^2,~\alpha^{\chi_{c2}}_\theta=\frac{|\mathcal{A}^{\chi_{c2}}_{2,1} |^2-2|\mathcal{A}^{\chi_{c2}}_{1 , 1} |^2+|\mathcal{A}^{\chi_{c2}}_{0,1} |^2}{|\mathcal{A}^{\chi_{c2}}_{2,1} |^2+2|\mathcal{A}^{\chi_{c2}}_{1 , 1} |^2+|\mathcal{A}^{\chi_{c2}}_{0,1} |^2}.
\end{eqnarray}

After integration over the polar angle $\theta$, the corresponding total cross sections are
\begin{eqnarray}
\label{sigma0}
&&\sigma(\eta_c+\gamma)=\frac{\alpha}{6s^2}\left(\frac{|\mathbf{P_1}|}{\sqrt{s} }\right)2|\mathcal{A}^{\eta_c}_{0 , 1}|^2,\nonumber\\
&&\sigma(\chi_{c0}+\gamma)=\frac{\alpha}{6s^2}\left(\frac{|\mathbf{P_1}|}{\sqrt{s} }\right)2|\mathcal{A}^{\chi_{c0}}_{0 , 1}|^2,\nonumber\\
&&\sigma(\chi_{c1}+\gamma)=\frac{\alpha}{6s^2}\left(\frac{|\mathbf{P_1}|}{\sqrt{s} }\right)(2|\mathcal{A}^{\chi_{c1}}_{1 , 1}|^2+2|\mathcal{A}^{\chi_{c1}}_{0 , 1}|^2),\nonumber\\
&&\sigma(\chi_{c2}+\gamma)=\frac{\alpha}{6s^2}\left(\frac{|\mathbf{P_1}|}{\sqrt{s} }\right)(2|\mathcal{A}^{\chi_{c2}}_{2 ,1}|^2+2|\mathcal{A}^{\chi_{c2}}_{1 , 1}|^2+2|\mathcal{A}^{\chi_{c2}}_{0 , 1}|^2).
\end{eqnarray}

\subsection{NRQCD factorization for the helicity amplitudes~\label{subsec:proj}}

The helicity amplitude $\mathcal{A}^{H}_{\lambda_1 , \lambda_2}$ can be evaluated using the NRQCD factorization: 
\begin{eqnarray}
\label{nrqcd-formula}
&&\mathcal{A}^{H}_{\lambda_1 , \lambda_2}=c^{H}_{\lambda_1 , \lambda_2} \sqrt{2M_H}\frac{\langle \mathcal{O}_H\rangle}{m_c^{n}},
\end{eqnarray}
where $n = 1$ for $\eta_c$ and $n=2$ for $\chi_{cJ}$, $c^{H}_{\lambda_1 , \lambda_2}$ denote the dimensionless short-distance coefficients (SDCs), and the NRQCD long-distance matrix elements (LDMEs) are defined as:
\begin{eqnarray}
\label{ld}
&&\langle\mathcal{O}_{\eta_c}\rangle=\langle \eta_c|\psi^\dagger\chi(\mu_\Lambda)|0\rangle,\nonumber\\
&&\langle\mathcal{O}_{\chi_{c0}}\rangle=\langle \chi_{c0}|\psi^\dagger\frac{1}{\sqrt{3}}\left (-\frac{i}{2}\overleftrightarrow{D}\cdot\sigma \right )\chi(\mu_\Lambda)|0\rangle,\nonumber\\
&&\langle\mathcal{O}_{\chi_{c1}}\rangle=\langle \chi_{c1}|\psi^\dagger\frac{1}{\sqrt{2}}\left (-\frac{i}{2}\overleftrightarrow{D}\times\sigma\right )\cdot\epsilon_{\chi_{c1}}\chi(\mu_\Lambda)|0\rangle,\nonumber\\
&&\langle\mathcal{O}_{\chi_{c2}}\rangle=\langle \chi_{c2}|\psi^\dagger\left( -\frac{i}{2}\overleftrightarrow{D}^{(i}\sigma^{j)}\epsilon^{ij}_{\chi{c2}}\right)\chi(\mu_\Lambda)|0\rangle,
\end{eqnarray}
with $\epsilon_{\chi_{c1}}$ and $\epsilon_{\chi_{c2}}$ denoting the polarization vector of $\chi_{c1}$ and the polarization tensor of $\chi_{c2}$, respectively. 
Here, the factorization scale $\mu_\Lambda$ is expected to cancel with the corresponding scale dependence in the SDCs. In Eq.~(\ref{nrqcd-formula}), the factor $\sqrt{2M_H}$ appears on the right side because
we adopt relativistic normalization for the charmonium state $H$, but we use conventional nonrelativistic normalization for the NRQCD LDMEs. In this work, we take $M_H=2 m_c$.

As the SDCs are insensitive to nonperturbative hadronization effects, they can be determined via the standard perturbative matching procedure. To this end, we replace the physical $H$ meson with a fictitious onium state consisting of a free $c\bar{c}$ pair carrying the same quantum numbers  $^{2S+1}L_J$ as $H$—such as $^1S_0$ for $\eta_c$ and $^3P_J$ for $\chi_{cJ}$. This enables us to compute both sides of Eq.~\eqref{nrqcd-formula} order by order in $\alpha_s$. 

Under this replacement, Eq.~\eqref{nrqcd-formula} becomes:
\begin{equation}
\mathcal{A}^{c\bar{c}(^{2S+1}L_J)}_{\lambda_1 , \lambda_2}= c^{H}_{\lambda_1 , \lambda_2} \frac{\langle \mathcal{O}_{c\bar{c}(^{2S+1}L_J)}\rangle}{m_c^{n}},
\label{nrqcd-formula-perp}
\end{equation}
where both sides are now accessible in perturbation theory. 
Note that the factor $\sqrt{2M_H}$ is omitted in Eq.~\eqref{nrqcd-formula-perp}, as we employ relativistic normalization for both the heavy quark states in the full QCD helicity amplitude and the NRQCD LDMEs. 
By evaluating both sides consistently in perturbative QCD and NRQCD, we can systematically extract the desired SDCs $c^{H}_{\lambda_1 , \lambda_2}$ as a power series in $\alpha_s$.

To facilitate the perturbative calculation, we assign the momenta of the $c$ and $\bar{c}$ quarks as
 \begin{subequations}\label{kinematics-momenta}
 \bqa
p&=&\frac{P}{2}+q, \\
\bar{p}&=&\frac{P}{2}-q,
\eqa
\end{subequations}
where $P$ and $q$ denote the total and relative momenta of the $c\bar{c}$ pair, respectively.
The on-shell condition $p^2=\bar{p}^2=m_c^2$ implies:
\bqa
\label{kinematics-on-shell}
&&P\cdot q=0,\qquad P^2=4E^2,
\eqa
with $E=\sqrt{m_c^2-q^2}\ge m_c$. Since we are only concerned with the SDCs at the lowest order in velocity $v$,
we may approximate the invariant mass of the $c\bar{c}$ pair by $4m_c^2$.

It is convenient to employ the covariant spin projectors to enforce the $c\bar{c}$ pair in a spin-singlet or spin-triplet state. The relativistically normalized color-singlet projectors for the spin-singlet and spin-triplet states are given by~\cite{Bodwin:2002cfe}:
\bqa
\label{spin-projector}
\Pi_0=\frac{1}{8\sqrt{2}m_c^2}(\pbarslash-m_c)\gamma^5(\Pslash+2m_c)(\pslash+m_c)\otimes {{\bf 1}_c \over \sqrt{N_c}}.\\
\Pi_1^\mu=\frac{-1}{8\sqrt{2}m_c^2}(\pbarslash-m_c)\gamma^\mu(\Pslash+2m_c)(\pslash+m_c)\otimes {{\bf 1}_c \over \sqrt{N_c}}.
\eqa

The amplitude for $\gamma^*\to c\bar{c}(^1S_0)+\gamma$ is obtained by replacing the product of charm spinors $v(\bar{p})\bar{u}(p)$ with $\Pi_0$:
 \bqa
\label{s-wave-projector}
\mathcal{A}^{(c\bar{c}(^1S_0))}={\rm Tr}[\mathcal{A}\Pi_0]\Big |_{q=0},
\eqa
where $\mathcal{A}$  is the quark-level amplitude for $\gamma^*\to c\bar{c}+\gamma$ with external quark spinors truncated. Since we work at the lowest order in velocity expansion, we set the relative momentum $q=0$.

For the process $\gamma^*\to c\bar{c}({}^3P_J)+\gamma$ , the amplitude is obtained by differentiating the color-singlet spin-triplet quark-level amplitude with respect to $q$ and then setting $q=0$:
 \bqa
\label{p-wave-projector}
\mathcal{A}^{(c\bar{c}({}^3P_J))}={\cal P}_J^{\mu\nu}\frac{d}{dq^\nu}{\rm Tr}[\mathcal{A} \Pi_{1,\mu}]\Big |_{q=0},
\eqa
where projectors ${\cal P}_J^{\mu\nu}$, which encode the Clebsch–Gordan coupling between the orbital and spin angular momenta, are defined as:
\begin{eqnarray}{\label{angularoperator}}
{\cal P}^{\mu\nu}_0&=&\frac{1}{\sqrt{3}}
(-g^{\mu\nu}+\frac{P^{\mu}P^{\nu}}{4m_c^2}),\nonumber\\
{\cal
P}^{\mu\nu}_1&=&\frac{i}{2\sqrt{2}m_c}
\epsilon^{\mu\nu\rho\sigma}P_\rho\epsilon^{*}_{\chi_{c1},\sigma},\nonumber\\
{\cal P}^{\mu\nu}_2&=&\epsilon^{*\mu\nu}_{\chi_{c2}}.
\end{eqnarray}

Using the helicity projection operators derived in Ref.~\cite{Zhang:2021ted}, we extract the perturbative helicity amplitude $\mathcal{A}^{c\bar{c}(^{2S+1}L_J)}_{\lambda_1, \lambda_2}$. The corresponding perturbative NRQCD matrix elements $\langle \mathcal{O}_{c\bar{c}(^{2S+1}L_J)} \rangle$ are then evaluated. Using both components, we can determine the SDCs $c^H_{\lambda_1 , \lambda_2}$ through the matching procedure.

\subsection{SDCs in $\alpha_s$ expansion}

Generally, the SDC $c^H_{\lambda_1 , \lambda_2}$ admits an expansion in powers of $\alpha_s$:
\begin{eqnarray}\label{eq-sdcs-in-as}
c^H_{\lambda_1 , \lambda_2}(r)&=&c^{H,(0)}_{\lambda_1 , \lambda_2}(r)
\Big[1+\frac{\alpha_s}{\pi}c^{H,(1)}_{\lambda_1 , \lambda_2}(r)+\frac{\alpha_s^2}{\pi^2} \Big(\frac{\beta_0}{4}c^{H,(1)}_{\lambda_1 , \lambda_2}(r)\ln \frac{4\mu_R^2}{s}\nonumber\\
&+&\gamma_H\ln \frac{\mu_\Lambda^2}{m_{c}^{2}}+c^{H,(2)}_{\lambda_1 , \lambda_2}(r)\Big)+\mathcal{O}\left(\alpha_s^3\right)\Big],
\end{eqnarray}
where $\beta_0 = \frac{11}{3} C_A - \frac{4}{3} T_f n_f$ is the one-loop coefficient of the QCD beta function, with $n_f = n_l + n_h$ denoting the number of active quark flavors. Here, $n_l = 3$ is the number of light quarks and $n_h = 1$ is the number of heavy quarks.  The scales $\mu_R$ and $\mu_\Lambda$ represent the renormalization scale and the NRQCD factorization scale, respectively.

The logarithmic term $\ln\mu_R$ ensures the renormalization group invariance of $c^H_{\lambda_1 , \lambda_2}(r)$ at two-loop accuracy, while the $\ln\mu_\Lambda$  term arises from the NRQCD factorization. In accordance with the factorization formalism, the $\ln\mu_\Lambda$-dependence in the SDCs must cancel exactly with that in the LDMEs.

The anomalous dimensions $\gamma_{\eta_c}$ and $\gamma_{\chi_{cJ}}$ associated with the NRQCD bilinear currents for the $^1S_0$ and $^3P_J$ states are given by~\cite{Czarnecki:1997vz,Beneke:1997jm,Sang:2015uxg,Hoang:2006ty,Sang:2020fql}:
\begin{eqnarray}\label{eq-sdcs}
&&\gamma_{\eta_c}=-\pi^2\left(\frac{C_A C_F}{4}+\frac{C_F^2}{2}\right),\nonumber\\
&&\gamma_{\chi_{c0}}=-\pi^2 \left(\frac{C_A C_F}{12}+\frac{C_{F}^{2}}{3}\right),\nonumber\\
&&\gamma_{\chi_{c1}}=-\pi^2 \left(\frac{C_A C_F}{12}+\frac{5C_{F}^{2}}{24}\right),\nonumber\\
&&\gamma_{\chi_{c2}}=-\pi^2 \left(\frac{C_A C_F}{12}+\frac{13C_{F}^{2}}{120}\right).
\end{eqnarray}

The LO SDCs $c^{H,(0)}_{\lambda_1 , \lambda_2}(r)$ were previously derived in Ref.~\cite{Chung:2008km}:
\begin{eqnarray}
\label{helicity0}
&&c^{\eta_c,(0)}_{0, 1}(r)=-\frac{16 \pi  \alpha }{9},\nonumber\\
&&c^{\chi_{c0},(0)}_{0,1}(r)=\frac{16 \pi  \alpha  (3r-1)}{9 \sqrt{3} (r-1)},\nonumber\\
&&c^{\chi_{c1},(0)}_{1,1}(r)=\frac{16 \sqrt{2} \pi  \alpha \sqrt{r}}{9(1-r)},\quad c^{\chi_{c1},(0)}_{0,1}(r)=\frac{16 \sqrt{2} \pi  \alpha}{9(1- r)},\nonumber\\
&&c^{\chi_{c2},(0)}_{2,1}(r)=\frac{32 \pi  \alpha  r}{9(r-1)},\quad
c^{\chi_{c2},(0)}_{1,1}(r)=\frac{16  \pi  \alpha \sqrt{2r}}{9(1- r)},\quad
c^{\chi_{c2},(0)}_{0,1}(r)=\sqrt{\frac{2}{3}}\frac{16 \pi  \alpha }{9 (r-1)},
\end{eqnarray}

Analytical expressions for the various NLO helicity SDCs can be found in Ref.~\cite{Sang:2009jc}, which also include their asymptotic behaviors in the limits $r\to 0$ and $r\to \infty$. Numerical results for these coefficients are provided in Ref.~\cite{Li:2009ki}.

At $\mathcal{O}(\alpha_s^2)$, the SDC for $e^+e^-\to \eta_c+\gamma$ has been computed analytically in terms of Goncharov polylogarithms, Chen's iterated integrals, and elliptic functions~\cite{Chen:2017pyi}. The correction was later revised numerically in Ref.~\cite{Yu:2020tri}. For $e^+e^-\to \chi_{cJ}+\gamma$, only the unpolarized cross sections have been studied numerically; the helicity amplitudes and polarized cross sections have not yet been calculated.

In this work, we compute the $c^H_{\lambda_1,\lambda_2}$ up to $\mathcal{O}(\alpha_s^2)$ and present them as asymptotic expansions around four discrete values of $r$.

\section{Computational technologies~\label{sec:com:tec}}

\begin{figure}[h!]
\begin{center}
\includegraphics[width=0.80\textwidth]{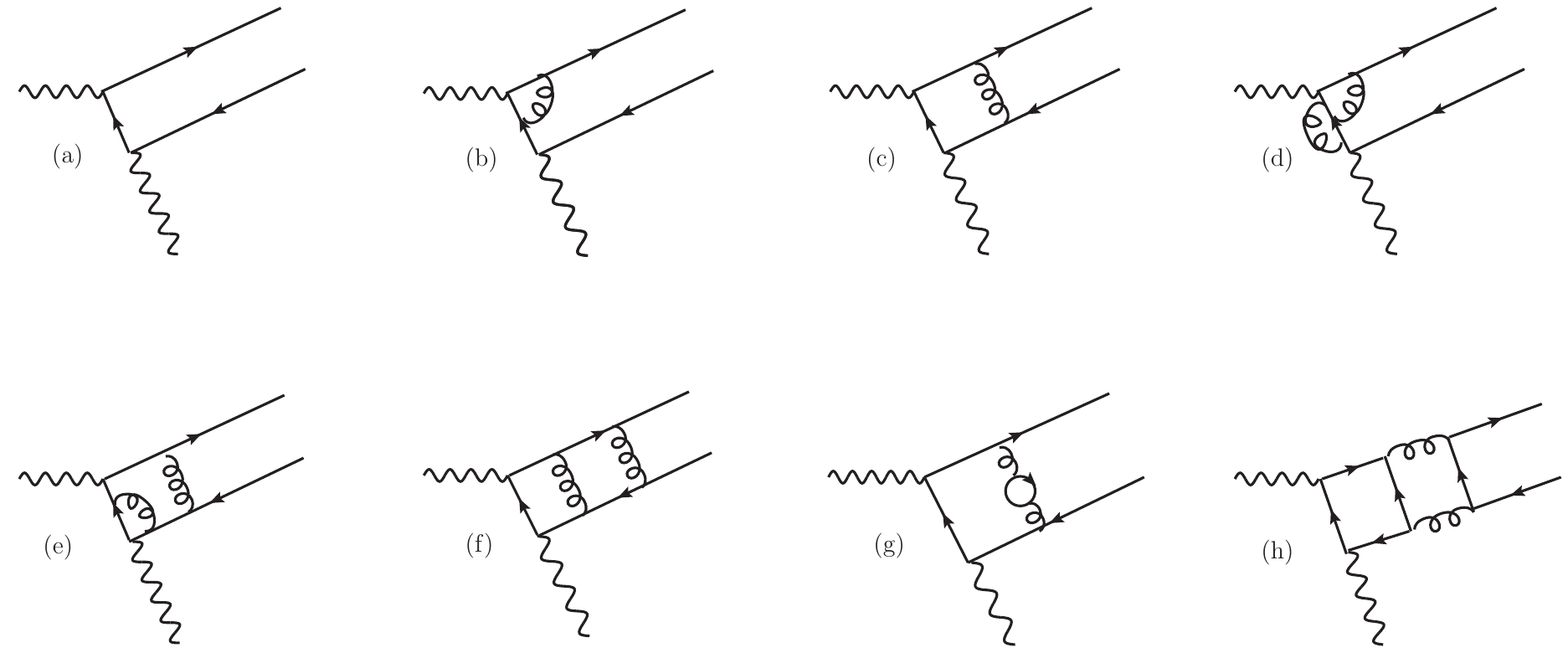}
\end{center}
\caption{Representative Feynman diagrams for the process $\gamma^* \to c\bar{c}+\gamma$ at various perturbative orders. (a) LO diagram; (b)–(c) NLO diagrams; (d)–(h) NNLO diagrams, where (h) corresponds to the light-by-light contribution. All diagrams were drawn using {\tt JaxoDraw}~\cite{Binosi:2008ig}.}
\label{feynman}
\end{figure}

We employ FeynArts~\cite{Hahn:2000kx} to generate the Feynman diagrams and corresponding amplitudes for the process $\gamma^{*} \to c \bar{c} + \gamma$ up to $\mathcal{O}(\alpha_s)$, with representative diagrams illustrated in Fig.~\ref{feynman}. 

Following the method outlined in Sec.~\ref{subsec:proj},
we apply spin/color, orbital-spin coupling, and helicity projectors to extract the relevant perturbative helicity amplitudes. The symbolic manipulation of Lorentz contractions and Dirac algebra is carried out using {\tt FeynCalc} and {\tt FormLink}~\cite{Mertig:1990an,Feng:2012tk}. 

We adopt the hard-region strategy~\cite{Beneke:1997zp} to directly extract the SDCs. Feynman integrals are categorized using {\tt CalcLoop}~\cite{CalcLoop}.
For the reduction of scalar integrals to master integrals (MIs), we utilize both {\tt Kira}~\cite{Klappert:2020nbg} and  {\tt Blade}~\cite{Guan:2024byi}, cross-checking the results to ensure consistency. We further apply {\tt FIRE}~\cite{Smirnov:2014hma} to identify symmetries among MIs from different integral families.  Ultimately, we obtain approximately 17 families and 150 master integrals. All calculations are performed in $d=4-2\epsilon$ spacetime dimensions to regularize ultraviolet and infrared (IR) divergences.

The asymptotic expansions of the MIs around $r=r_0$ are obtained by solving their corresponding differential equations (DEs) with respect to $r$. For a complete set of MIs $\vec{J}$, the DEs read:
\bqa
\frac{\partial }{\partial r}\vec{J}=M(r,\epsilon) \vec{J},
\label{eq-de}
\eqa
where the matrix $M$ takes a local Fuchsian form at $r=r_0$. The solution for $\vec{J}$ admits a Laurent expansion in $r-r_0$:

\bqa
\vec{J}=\sum_{\rho \in S} (r-r_0)^\rho \sum_{n=0}^{\infty}\vec{C}_{\rho,n}(\epsilon) (r-r_0)^n,
\label{eq-asy}
\eqa
with $S$ being a finite set of 
$\epsilon$-dependent exponents. Substituting~\eqref{eq-asy} into~\eqref{eq-de}, we determine the exponents $\rho$ by matching the lowest-order terms and derive the coefficients $\vec{C}$ from recurrence relations and initial conditions.

In practice, we adopt a numerical fitting strategy: we first solve the DEs for a set of discrete $\epsilon$ values ($\epsilon_0, ..., \epsilon_n$). Then, by solving a system of linear equations, we determine $\rho$ and $\vec{C}$ as expansions in $\epsilon$
\bqa
\vec{a}_0+\vec{a}_1\epsilon+...+\vec{a}_n\epsilon^n.
\eqa
This method proves highly efficient and is implemented using the auxiliary mass flow (AMFlow) package~\cite{Liu:2017jxz,Liu:2022mfb,Liu:2022chg}.

Once the MIs are determined, we further expand both $\rho$ and $\vec{C}$ in $\epsilon$. Each integral is then expressed as a series in $\epsilon$ and $r-r_0$, yielding final results of the form $\sum_{i,j}  f^{H,(n)}_{\lambda_1,\lambda_2}(i,j)(r-r_0)^i\ln^j (r-r_0)$ at each order in $\epsilon$.  Thanks to the high precision of the coefficients $f^{H,(n)}_{\lambda_1,\lambda_2}(i,j)$,  we successfully reconstruct their analytic expressions up to $\ln^2 (r-r_0)$ for lower powers of $r-r_0$ using the {\tt PSLQ} algorithm. For related treatments, we refer interested readers to Refs.~\cite{Li:2025mng,Chen:2025qgy,Chen:2025ocd}.

In this study, we adopt the on-shell scheme to renormalize the charm quark mass and field strength, and the $\overline{\text{MS}}$ scheme for the QCD coupling constant. The renormalized NNLO amplitude contains a single IR pole, whose coefficient exactly equals half of the anomalous dimension of the NRQCD bilinear operator with the quantum number of $\eta_c$ or $\chi_{cJ}$~\cite{Czarnecki:1997vz,Beneke:1997jm,Sang:2015uxg,Hoang:2006ty,Sang:2020fql}. In the $\overline{\text{MS}}$ scheme, this IR pole factorizes into the corresponding NRQCD long-distance matrix elements.

\section{Asymptotic expressions of the SDCs around $r=0$~\label{sec:asy:exp}}

After extensive calculations, we have obtained asymptotic expansions of $c^{H,(n)}_{\lambda_1,\lambda_2}(r)$ (with $n=1, 2$) around the four values $r_0$: $0$, $1/3$, $2/3$, and $1$.
The coefficients are expressed as series of the form
\begin{equation}
c^{H,(n)}_{\lambda_1,\lambda_2}(r) = \sum_{j\ge 0} f^{H,(n)}_{\lambda_1,\lambda_2}(i,j)(r - r_0)^i \ln^j (r - r_0),
\end{equation}
where the power $i$ is truncated at $30$ for $r_0=0$ and at $15$ for the other expanding points. Note that the expansions at $r_0=1/3$ and $r_0=2/3$ contain no logarithmic terms. In what follows we focus on the expansion around $r_0 = 0$.

The leading-logarithmic coefficients at NLO read:
\begin{eqnarray}\label{exp:1loop:ll}
f^{\eta_c,(1)}_{0,1}(0,1)&=&\left(\frac{\ln 2}{2}-\frac{3}{4}\right)C_F,\nonumber \\
f^{\chi_{c0},(1)}_{0,1}(0,1)&=&\left(\frac{\ln 2}{2}-\frac{1}{4}\right)C_F,\nonumber \\
f^{\chi_{c1},(1)}_{1,1}(0,1)&=&\left(\frac{\ln 2}{2}-\frac{3}{4}\right)C_F,\quad
f^{\chi_{c1},(1)}_{0,1}(0,1)=\left(\frac{\ln 2}{2}-\frac{3}{4}\right)C_F,\nonumber \\
f^{\chi_{c2},(1)}_{2,1}(0,1)&=&(2\ln 2 -1) C_F,\quad
f^{\chi_{c2},(1)}_{1,1}(0,1)=\left(\frac{1}{4}+\frac{\ln 2}{2}\right) C_F,\nonumber\\
f^{\chi_{c2},(1)}_{0,1}(0,1)&=&\left(\frac{\ln 2}{2}-\frac{1}{4}\right)C_F.
\end{eqnarray}
which agree with the results given in Refs.~\cite{Jia:2008ep,Sang:2009jc,Wang:2013ywc}.

At NNLO, we observe the appearance of ``superleading" terms in the expansion of the helicity-suppressed coefficient $c^{\chi_{c2},(n)}_{2,1}(r)$:
\begin{eqnarray}\label{exp:2loop:superleading}
f^{\chi_{c2},(2)}_{2,1}(-1,0)&=&\left(\frac{5}{4}\ln 2\, C_FT_f\right)_{\rm lbl}, \quad 
f^{\chi_{c2},(2)}_{2,1}(0,3)=\left(-\frac{1}{24}\,C_F T_f\right)_{\rm lbl},
\end{eqnarray}
where the subscript ``lbl" marks contributions originating from the light‑by‑light Feynman diagrams shown in Fig.~\ref{feynman}. No such ``superleading" terms appear in other helicity configurations.

The leading-logarithmic coefficients at NNLO are:
\begin{eqnarray}\label{one-ex1}
f^{\eta_c,(2)}_{0,1}(0,2)&=&\left(\frac{11}{32}-\frac{11\ln 2 }{48}\right) C_AC_F+\left(\frac{9}{32}-\frac{\pi ^2}{96}+\frac{\ln^2 2}{16}-\frac{\ln 2}{2}\right) C_F^2\nonumber \\
&+&\left(\frac{\ln2}{3}-\frac{1}{2}\right)C_F T_f+\left[\left(-\frac{5 \pi^2}{48}-\frac{5 \ln^2 2}{8}+\frac{15 \ln2}{8}\right)C_FT_f\right]_{\rm lbl} ,\nonumber\\
f^{\chi_{c0},(2)}_{0,1}(0,2)&=&\left(\frac{11}{96}-\frac{11\ln 2}{48}\right) C_AC_F
+\left(-\frac{7}{32}-\frac{\pi ^2}{96}+\frac{\ln^2 2}{16}+\frac{\ln 2}{8}\right) C_F^2\nonumber \\
&+&\left(\frac{\ln2}{3}-\frac{1}{6}\right)C_F T_f+\left[\left(\frac{5}{8}-\frac{5\pi ^2}{48}-\frac{5 \ln^2 2}{8}+\frac{15 \ln2}{8}\right) C_FT_f\right]_{\rm lbl},\nonumber\\
f^{\chi_{c1},(2)}_{1,1}(0,2)&=&\left(\frac{9}{32}-\frac{5\ln 2}{48}\right) C_AC_F
+\left(\frac{9}{32}-\frac{\pi ^2}{96}+\frac{\ln^2 2}{16}-\frac{\ln 2}{2}\right) C_F^2\nonumber \\
&+&\left(\frac{\ln2}{3}-\frac{1}{2}\right)C_F T_f+\left[\left(\frac{5 \pi^2}{24}+\frac{5 \ln^2 2}{4}-\frac{15 \ln2}{4}\right) C_FT_f\right]_{\rm lbl},\nonumber\\
f^{\chi_{c1},(2)}_{0,1}(0,2)&=&\left(\frac{11}{32}-\frac{11\ln 2}{48}\right) C_AC_F
+\left(\frac{9}{32}-\frac{\pi ^2}{96}+\frac{\ln^2 2}{16}-\frac{\ln 2}{2}\right) C_F^2\nonumber \\
&+&\left(\frac{\ln2}{3}-\frac{1}{2}\right)C_F T_f+\left[\left(-\frac{5 \pi^2}{48}-\frac{5 \ln^2 2}{8}+\frac{15 \ln2}{8}\right) C_FT_f\right]_{\rm lbl},\nonumber\\
f^{\chi_{c2},(2)}_{2,1}(0,2)&=&\left(\frac{1}{3}-\frac{2 \ln 2}{3}\right) C_AC_F
+\left(\frac{3}{8}+\frac{\pi ^2}{48}+\frac{5 \ln^2 2}{8}-\frac{11 \ln 2}{8}\right) C_F^2\nonumber \\
&+&\left(\frac{4 \ln2}{3}-\frac{2}{3}\right)C_F T_f+\left[\left(2-\frac{5 \pi^2}{48}-\frac{5 \ln^2 2}{8}+\frac{\ln2}{4}-\frac{i}{8} \pi   \right) C_FT_f\right]_{\rm lbl},\nonumber\\
f^{\chi_{c2},(2)}_{1,1}(0,2)&=&\left(-\frac{5}{96}-\frac{17\ln 2}{48}\right) C_AC_F
+\left(\frac{1}{32}-\frac{\pi ^2}{96}+\frac{\ln^2 2}{16}+\frac{\ln 2}{4}\right) C_F^2\nonumber \\
&&+\left(\frac{1}{6}+\frac{\ln2}{3}\right) C_FT_f+\left[\left(-\frac{5}{2}+\frac{5\pi ^2}{24}+\frac{5 \ln^2 2}{4}\right) C_FT_f\right]_{\rm lbl},\nonumber\\
f^{\chi_{c2},(2)}_{0,1}(0,2)&=&\left(\frac{11}{96}-\frac{11\ln 2}{48}\right) C_AC_F+\left(-\frac{7}{32}-\frac{\pi ^2}{96}+\frac{\ln^2 2}{16}+\frac{\ln 2}{8}\right) C_F^2\nonumber\\
&+&\left(\frac{\ln2}{3}-\frac{1}{6}\right)C_F T_f+\left[\left(\frac{5}{8}-\frac{5\pi ^2}{48}-\frac{5 \ln^2 2}{8}+\frac{15 \ln2}{8}\right) C_FT_f\right]_{\rm lbl}.
\end{eqnarray}
The coefficients $f^{\eta_{c},(2)}_{0,1}(0,2)$, $f^{\chi_{c0},(2)}_{0,1}(0,2)$, $f^{\chi_{c1},(2)}_{0,1}(0,2)$ and $f^{\chi_{c2},(2)}_{0,1}(0,2)$ associated with the leading helicity amplitudes agree with the results obtained from light-cone evolution in Refs.~\cite{Jia:2008ep,Sang:2023hjl,Sang:2022erv}.

The explicit expressions for the coefficients $f^{H,(1)}_{\lambda_1,\lambda_2}(i,j)$ with $i=0, j=0$ and $i=1, j=0, 1, 2$ are collected in Appendix~\ref{app:1loop}, and the corresponding coefficients $f^{H,(2)}_{\lambda_1,\lambda_2}(i,j)$ are given in Appendix~\ref{app:2loop}. In particular, the coefficient $f^{\eta_{c},(2)}_{0,1}(0,1)$ reproduces the analytic expression derived from light-cone evolution in Ref.~\cite{Sang:2023hjl}.

For the reader’s convenience we provide, as supplementary material, all expansion coefficients up to $r^{30}$ for $r_0=0$ and up to $15$ for $r_0=1/3, 2/3, 1$, together with the numerical results for $c^{H,(2)}_{\lambda_1,\lambda_2}$ at $2494$ values of $r$ ranging from $0$ to $1$. The numerical results are obtained by solving the DEs in $r$ with the {\tt AMFlow} package~\cite{Liu:2022chg}. 

\section{Confronting the asymptotic expressions with the exact results~\label{sec:conv}}
\subsection{Convergence behavior of the asymptotic expansion at $r=0$}

We examine the convergence of the asymptotic expansions in $r$ for the coefficients $c^{H,(1)}_{\lambda_1,\lambda_2}(r)$ and $c^{H,(2)}_{\lambda_1,\lambda_2}(r)$.
 To facilitate the analysis, we introduce the compact notation:
\begin{eqnarray}
&&g^{H,(n)}_{\lambda_1,\lambda_2}(a,b) = \sum_{j=-1}^{a}\sum_{k=b}^{4}f^{H,(n)}_{\lambda_1,\lambda_2}(j,k)\, r^j\ln^{k}r,
\end{eqnarray}
and denote by $g^{H,(n)}_{\lambda_1,\lambda_2;\rm tot}$ the full contribution of  $c^{H,(n)}_{\lambda_1,\lambda_2}(r)$.

\begin{figure*}[!h]
\begin{center}
\hspace{0cm}\includegraphics[width=0.495\textwidth]{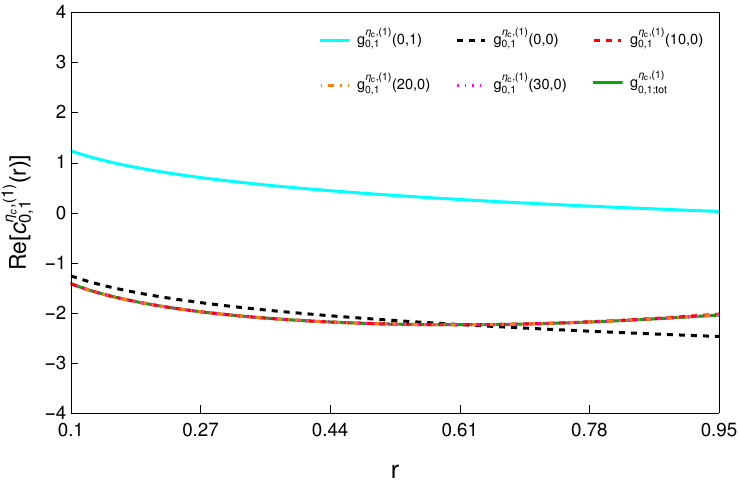}
\hspace{0cm}\includegraphics[width=0.495\textwidth]{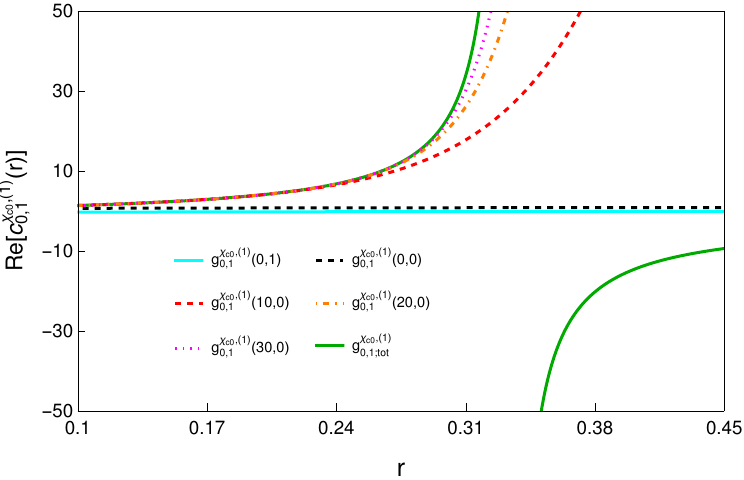}
\hspace{0cm}\includegraphics[width=0.495\textwidth]{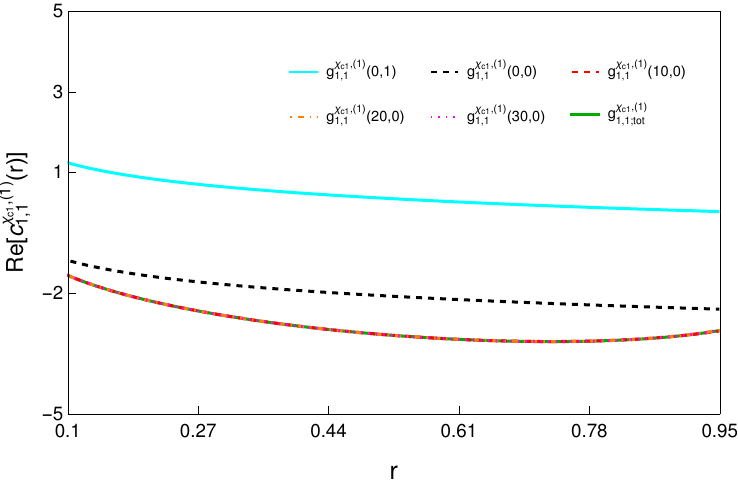}
\hspace{0cm}\includegraphics[width=0.495\textwidth]{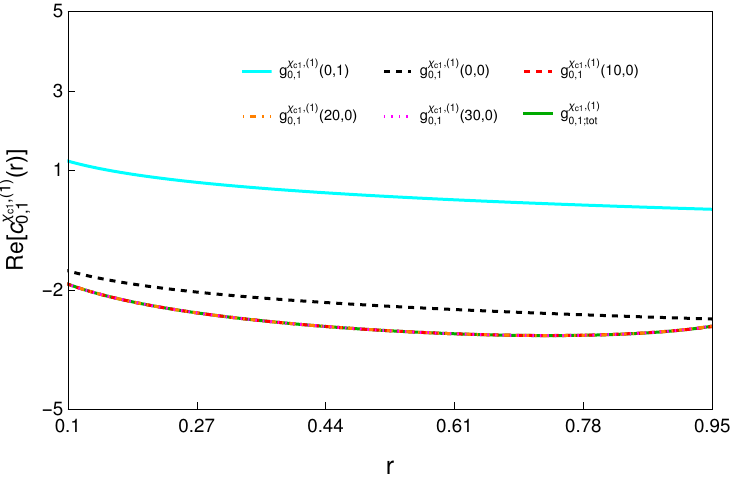}
\hspace{0cm}\includegraphics[width=0.495\textwidth]{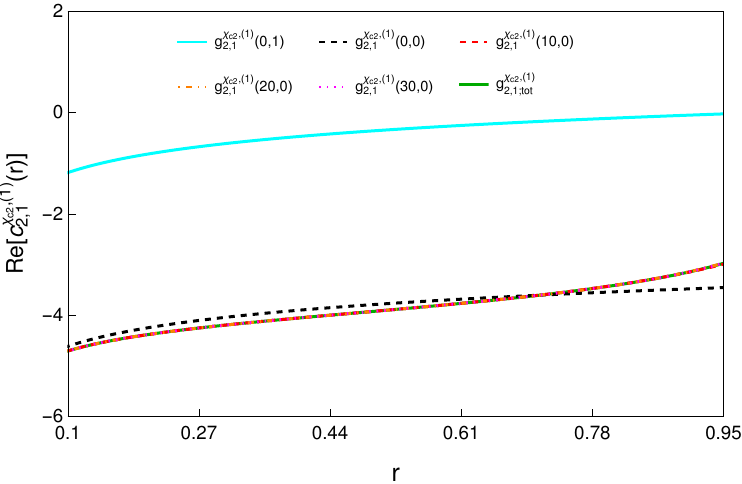}
\hspace{0cm}\includegraphics[width=0.495\textwidth]{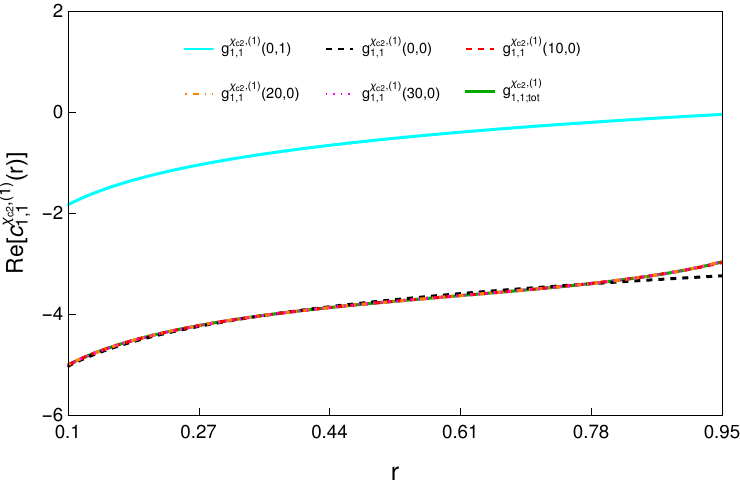}
\hspace{0cm}\includegraphics[width=0.495\textwidth]{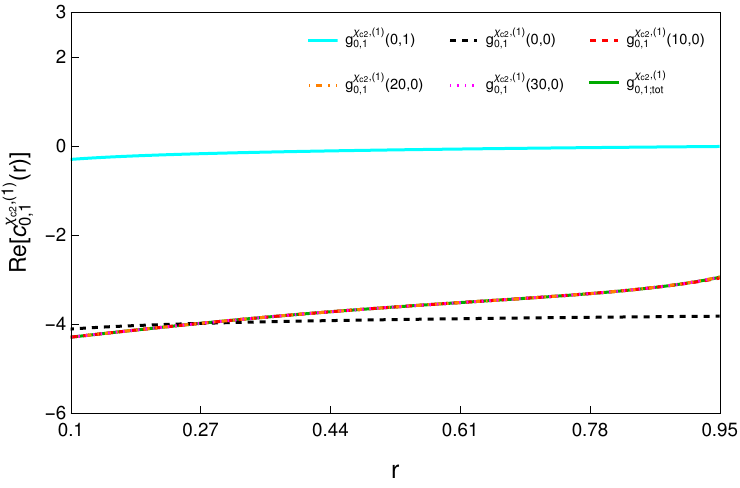}
\caption{\label{fig:loop1re}
Comparison between the asymptotic expansions around $r=0$ and the exact results for the real parts of the NLO coefficients $c^{H,(1)}_{\lambda_1,\lambda_2}$.}
\end{center}
\end{figure*}

Fig.~\ref{fig:loop1re} compares the real part of $c^{H,(1)}_{\lambda_1,\lambda_2}$ obtained from the asymptotic expansions with the exact results. While the leading-logarithmic term $r^0\ln r$ alone shows a sizable deviation from the exact value, the inclusion of higher-order terms in $r$ rapidly improves the convergence. Ultimately, the asymptotic expansions yield a reliable approximation across the entire range of $r$. 

An exception is $c^{\chi_{c0},(1)}_{0, 1}$, for which the asymptotic expression deviates markedly from the exact result for $r>0.3$. This behavior is attributed to a singularity in the exact result at $r=1/3$, which limits the convergence of the $r$-expansion  regardless of the number of terms included.

\begin{figure*}[!h]
\begin{center}
\hspace{0cm}\includegraphics[width=0.495\textwidth]{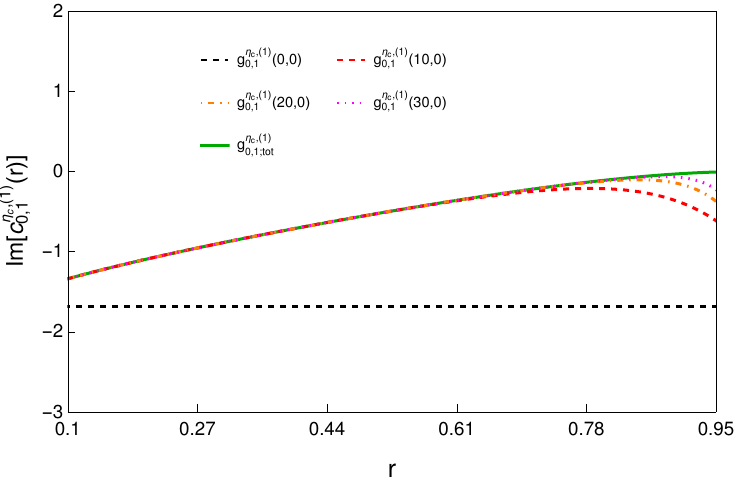}
\hspace{0cm}\includegraphics[width=0.495\textwidth]{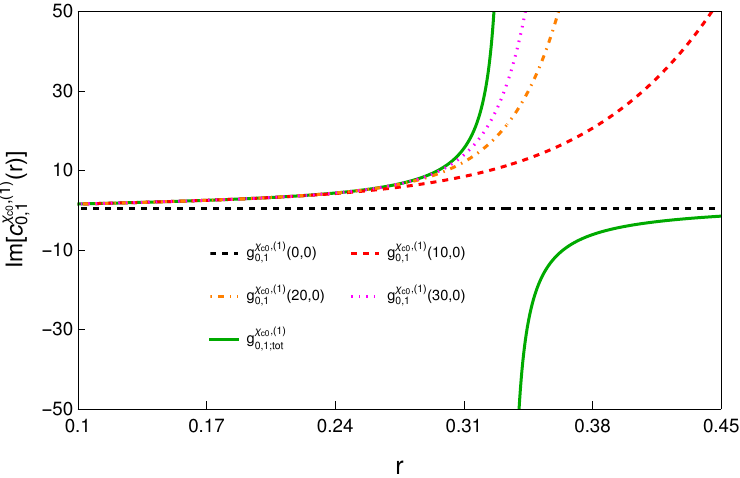}
\hspace{0cm}\includegraphics[width=0.495\textwidth]{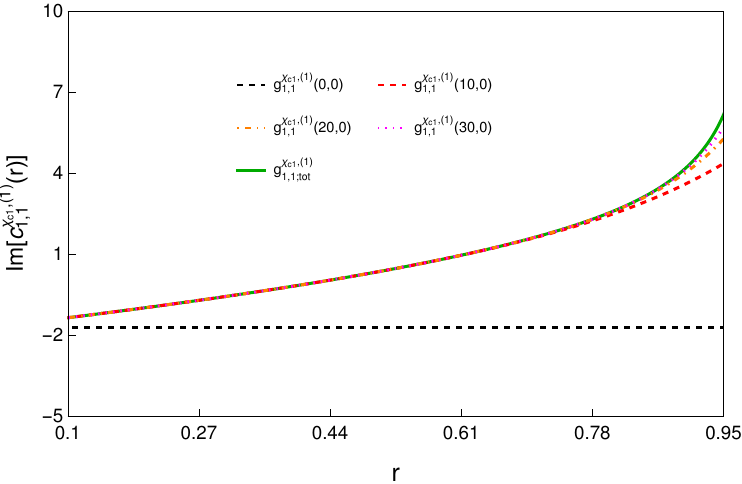}
\hspace{0cm}\includegraphics[width=0.495\textwidth]{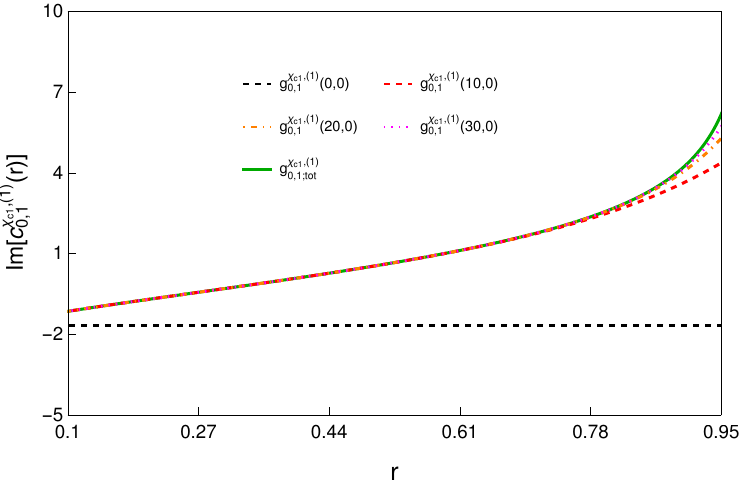}
\hspace{0cm}\includegraphics[width=0.495\textwidth]{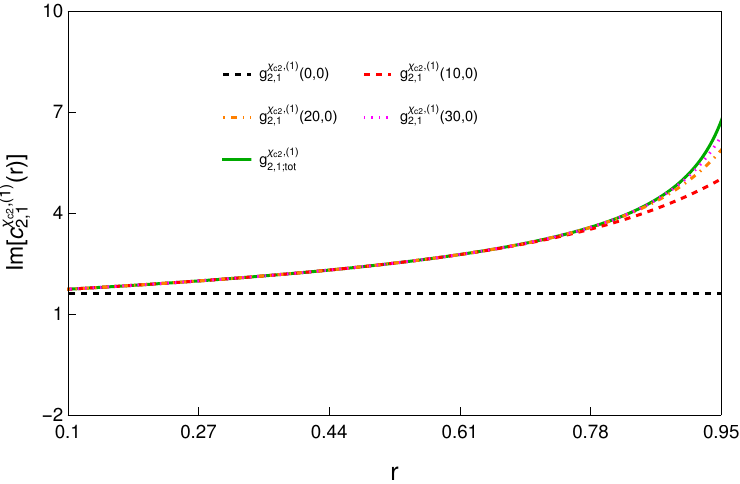}
\hspace{0cm}\includegraphics[width=0.495\textwidth]{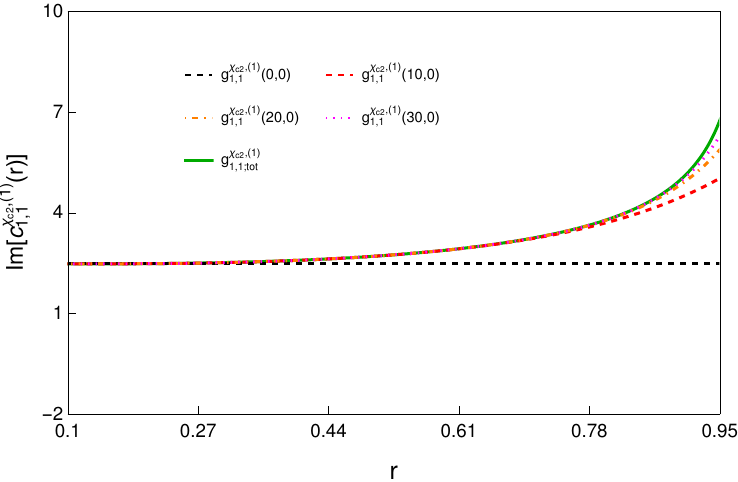}
\hspace{0cm}\includegraphics[width=0.495\textwidth]{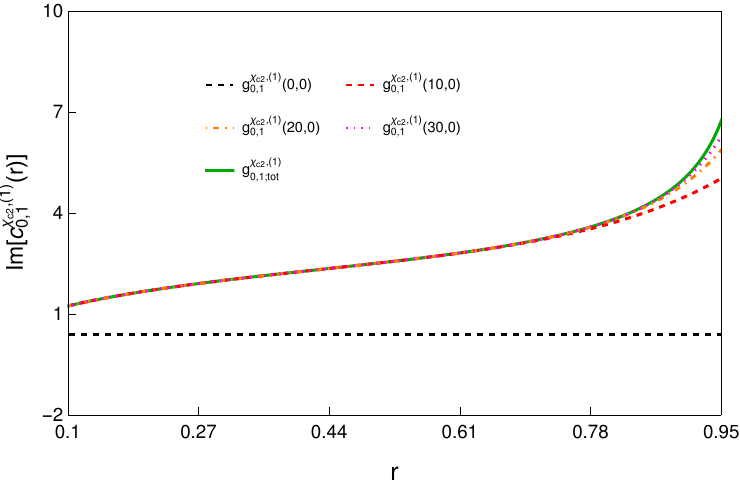}
\caption{\label{fig:loop1im}
Comparison between the asymptotic expansions around $r=0$ and the exact results for the imaginary parts of the NLO coefficients $c^{H,(1)}_{\lambda_1,\lambda_2}$.}
\end{center}
\end{figure*}

As shown in Fig.~\ref{fig:loop1im}, a similar comparison is performed for the imaginary part of $c^{H,(1)}_{\lambda_1,\lambda_2}$. In analogy to the real part, the asymptotic expansion agrees reasonably well with the exact results over most of the $r$-range, except again for $c^{\chi_{c0},(1)}_{0, 1}$. In contrast to the real part, however, convergence deteriorates as $r\to 1$. For example, when expanding up to $r^{30}$, the deviation reaches $4\%$ at $r = 0.82$ for $c^{\eta_c,(1)}_{0,1}$, and attains a similar level at $r=0.92$ for other coefficients.

\begin{figure*}[!h]
\begin{center}
\hspace{0cm}\includegraphics[width=0.495\textwidth]{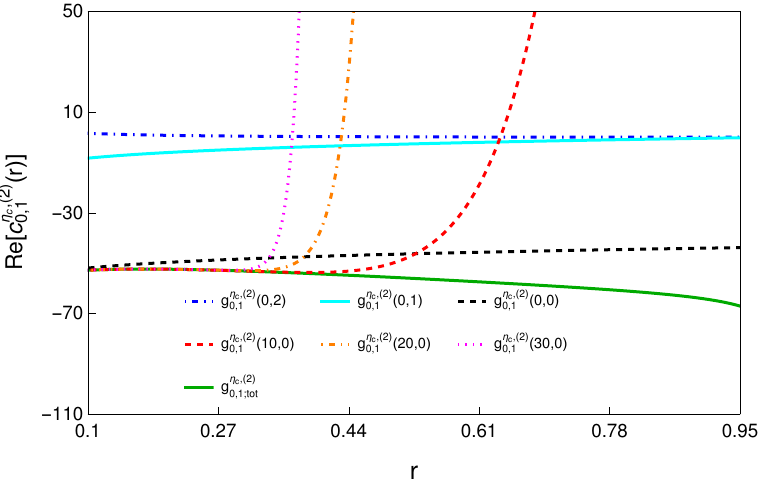}
\hspace{0cm}\includegraphics[width=0.495\textwidth]{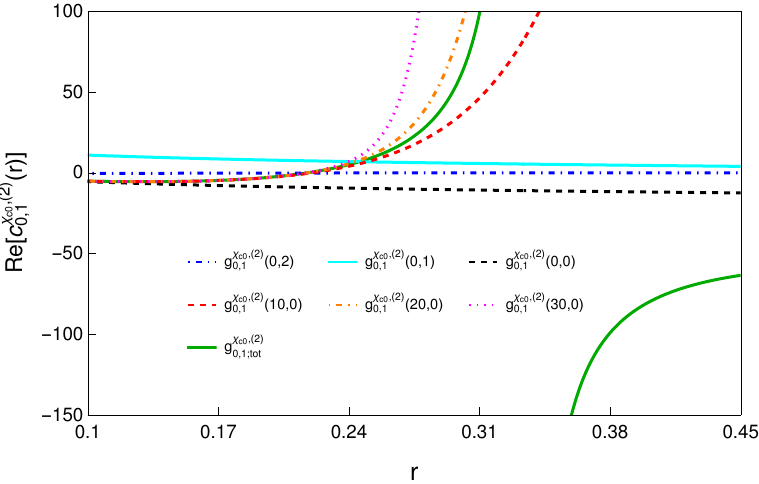}
\hspace{0cm}\includegraphics[width=0.495\textwidth]{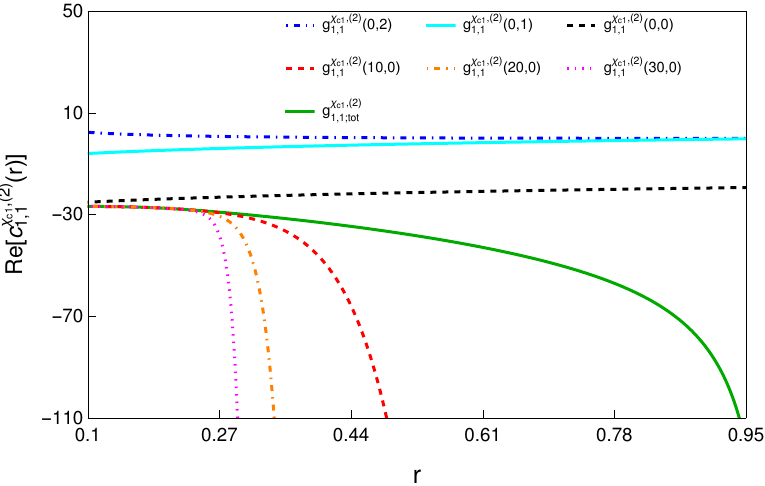}
\hspace{0cm}\includegraphics[width=0.495\textwidth]{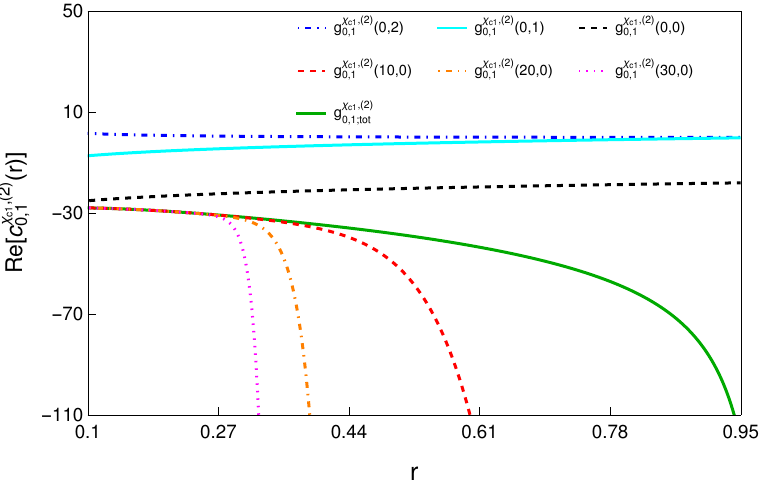}
\hspace{0cm}\includegraphics[width=0.495\textwidth]{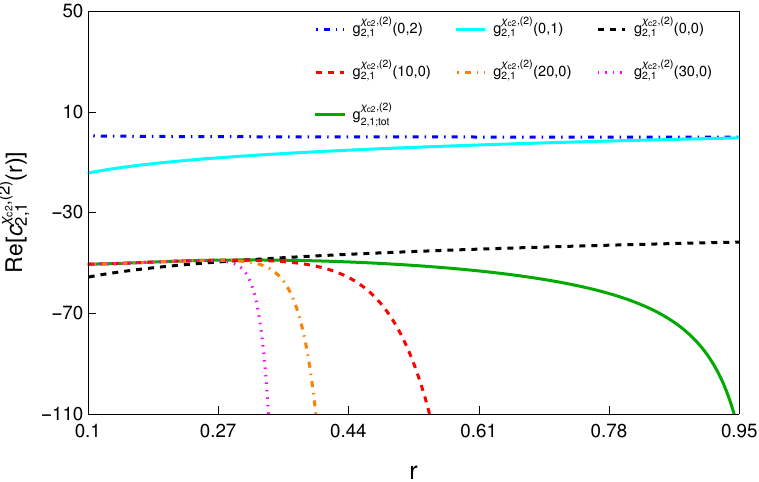}
\hspace{0cm}\includegraphics[width=0.495\textwidth]{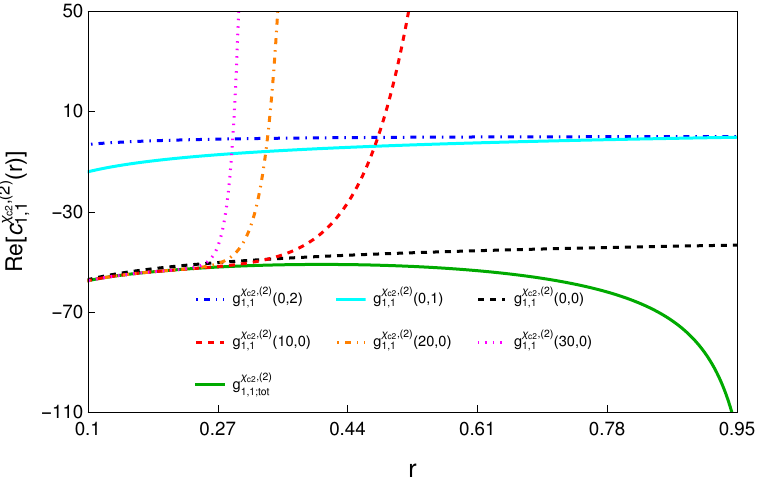}
\hspace{0cm}\includegraphics[width=0.495\textwidth]{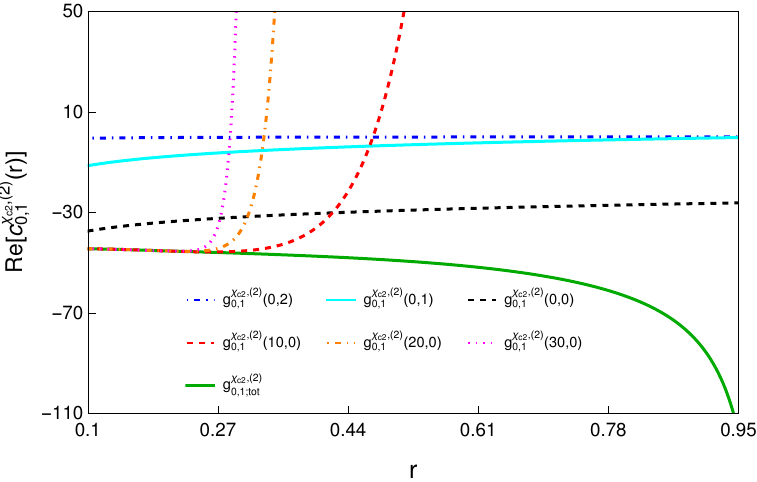}
\caption{\label{fig:loop2re}
Comparison between the asymptotic expansions around $r=0$ and the exact results for the real parts of the NNLO coefficients $c^{H,(2)}_{\lambda_1,\lambda_2}$.}
\end{center}
\end{figure*}

\begin{figure*}[!h]
\begin{center}
\hspace{0cm}\includegraphics[width=0.495\textwidth]{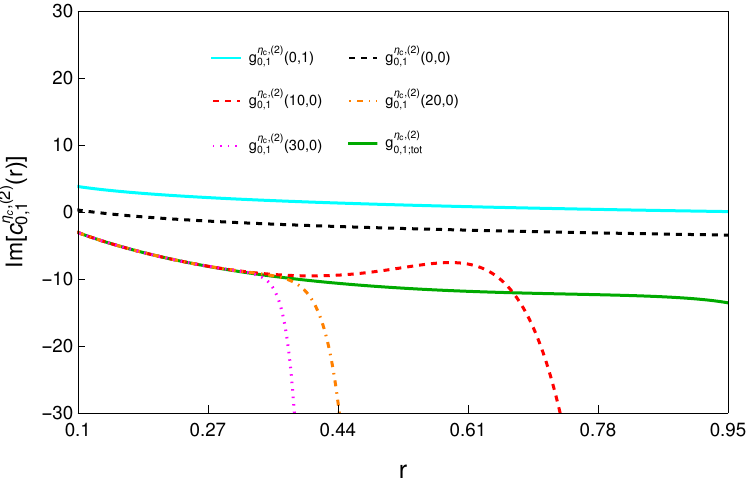}
\hspace{0cm}\includegraphics[width=0.495\textwidth]{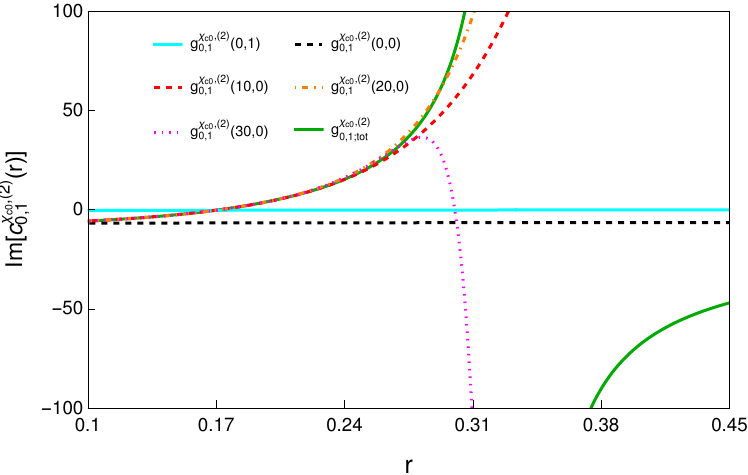}
\hspace{0cm}\includegraphics[width=0.495\textwidth]{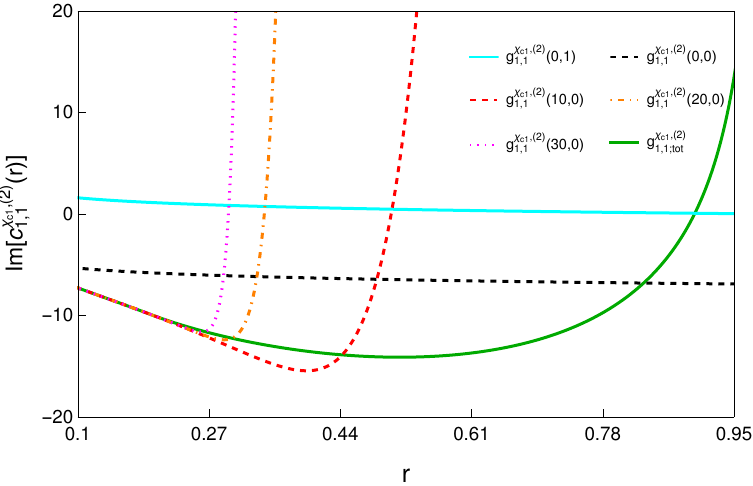}
\hspace{0cm}\includegraphics[width=0.495\textwidth]{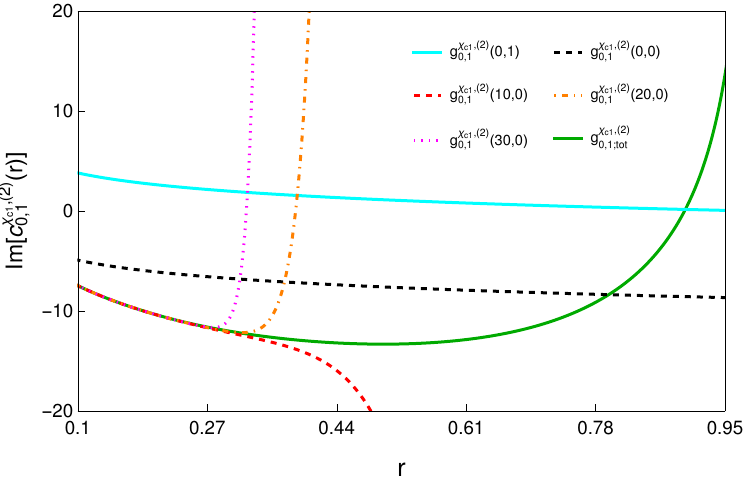}
\hspace{0cm}\includegraphics[width=0.495\textwidth]{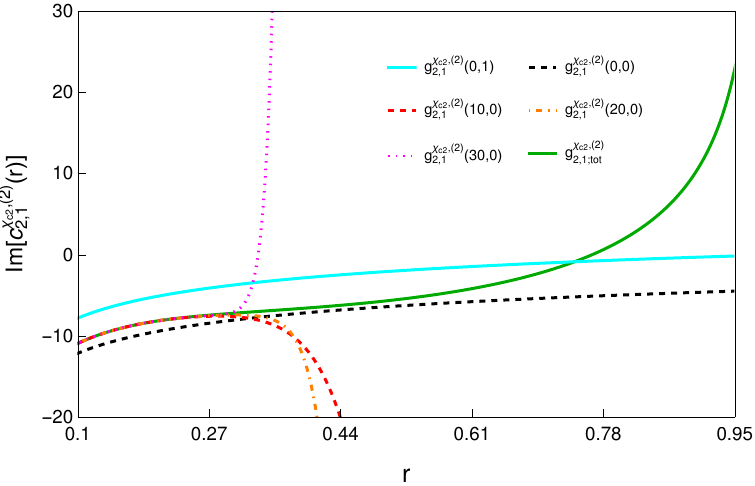}
\hspace{0cm}\includegraphics[width=0.495\textwidth]{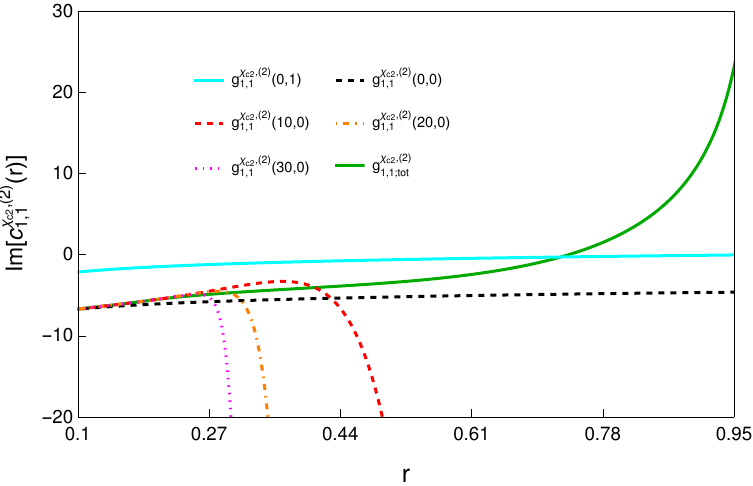}
\hspace{0cm}\includegraphics[width=0.495\textwidth]{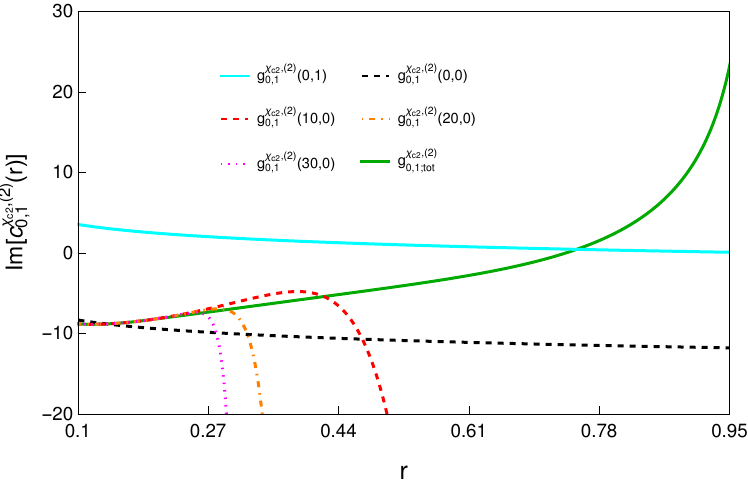}
\caption{\label{fig:loop2im}
Comparison between the asymptotic expansions around $r=0$ and the exact results for the imaginary parts of the NNLO coefficients $c^{H,(2)}_{\lambda_1,\lambda_2}$.}
\end{center}
\end{figure*}

We present in Fig.~\ref{fig:loop2re} and Fig.~\ref{fig:loop2im} a comparison of the real and imaginary parts, respectively, of $c^{H,(2)}_{\lambda_1,\lambda_2}$ between the asymptotic expansions and the exact results. It is observed that the asymptotic expressions reproduce the exact results accurately only at small $r$, while the deviation grows substantially as $r$ increases, for both the real and imaginary parts. 

For instance, at $r=0.27$, including terms up to $r^{30}$ keeps the deviation below $1\%$ for $c^{\eta_c,(2)}_{0, 1}$, in both real and imaginary components. At larger $r$ values, however, the deviation deteriorates noticeably. More strikingly, adding further terms in the
$r$-expansion beyond a certain point actually increases the discrepancy$-$a clear sign that the series no longer converges in this region. A similar loss of convergence already occurs at even lower 
$r$ for the other coefficients.

It is important to emphasize that despite the poor convergence at large $r$ values for $c^{H,(2)}_{\lambda_1,\lambda_2}$, the expansion performs excellently at smaller $r$. For example, at $r=0.2$, which corresponds to $\sqrt{s}=6.7$ GeV for $m_c=1.5$ GeV, the approximation remains highly accurate. Thus, the asymptotic expansion around $r_0=0$ can be safely applied to phenomenological predictions for $H+\gamma$ production at B factories.

\subsection{Combined asymptotic expansions around the four expansion points}

We investigate the asymptotic expansions of the NNLO coefficients $c^{H,(2)}_{\lambda_1,\lambda_2}(r)$ around the four points $r_0=0, 1/3, 2/3, 1$.

First, it is instructive to analyse the singular behaviours of these coefficients as $r\to r_0$.
At $r_0=0$ the asymptotic structure exhibits $1/r$ and $\ln^3 r$ singularities in $c^{\chi_{c2},(2)}_{2, 1}$, as shown in Eq.~(\ref{exp:2loop:superleading}), while only $\ln^2 r$ terms appear in the other coefficients.  Near $r_0=1/3$, a simple pole $1/(r-1/3)$ occurs in $c^{\chi_{c0},(2)}_{0,1}$, with no singular behavior observed in other coefficients. At $r_0\to 1$, only a logarithmic singularity $\ln (1-r)$ appears in $c^{\eta_c,(2)}_{0,1}$, whereas the other coefficients develop multiple singularities, including $1/(1-r), 1/\sqrt{1-r}, \ln(1-r)$.
Note that no singularities are present at $r_0=2/3$. A summary of all singular behaviors as $r\to r_0$ is provided in Table~\ref{tab:singular}. 

\begin{table}[!h]
\begin{center}
\caption{Leading singular behaviors of the NNLO coefficients $c^{H,(2)}_{\lambda_1,\lambda_2}$ in the limit $r\to r_0$.}
\label{tab:singular}
\begin{tabular}{|c|c|c|c|c|}
\hline
{$r_0$} & $0$ & $\frac{1}{3}$ & $\frac{2}{3}$ & $1$    \\ \hline
$c^{\eta_c,(2)}_{0,1}$    &$[0.31-(0.02)_{\rm lbl}]\ln^2 r$  &$\textbackslash$   &$\textbackslash$  &$5.85\ln(1-r)$  \\ 
$c^{\chi_{c0},(2)}_{0,1}$ &$[-0.50+(0.40)_{\rm lbl}]\ln^2 r$  &$-\frac{9.41+10.75 i}{3r-1}$ +$\frac{(0.33+0.79 i)_{\rm lbl}}{3r-1}$   &$\textbackslash$  &$\frac{2.59}{r-1}-\frac{3.14 i\ln(1-r)}{\sqrt{1-r}}$  \\ 
$c^{\chi_{c1},(2)}_{1,1}$ &$[0.41+(0.04)_{\rm lbl}]\ln^2 r$  &$\textbackslash$   &$\textbackslash$  &$\frac{2.59}{r-1}-\frac{3.14 i\ln(1-r)}{\sqrt{1-r}}$  \\ 
$c^{\chi_{c1},(2)}_{0,1}$ &$[0.31-(0.02)_{\rm lbl}]\ln^2 r$  &$\textbackslash$   &$\textbackslash$  &$\frac{2.59}{r-1}-\frac{3.14 i\ln(1-r)}{\sqrt{1-r}}$  \\ 
$c^{\chi_{c2},(2)}_{2,1}$ &$(\frac{0.58}{r})_{\rm lbl}-(0.03)_{\rm lbl}\ln^3 r$  &$\textbackslash$   &$\textbackslash$  &$\frac{2.59}{r-1}-\frac{3.14 i\ln(1-r)}{\sqrt{1-r}}$  \\ 
$c^{\chi_{c2},(2)}_{1,1}$ &$[-0.69+(0.10)_{\rm lbl}]\ln^2 r$  &$\textbackslash$   &$\textbackslash$  &$\frac{2.59}{r-1}-\frac{3.14 i\ln(1-r)}{\sqrt{1-r}}$  \\ 
$c^{\chi_{c2},(2)}_{0,1}$ &$[-0.50+(0.40)_{\rm lbl}]\ln^2 r$  &$\textbackslash$   &$\textbackslash$  &$\frac{2.59}{r-1}-\frac{3.14 i\ln(1-r)}{\sqrt{1-r}}$  \\ \hline
\end{tabular}
\end{center}
\end{table}

\begin{figure*}[!h]
\begin{center}
\hspace{0cm}\includegraphics[width=0.495\textwidth]{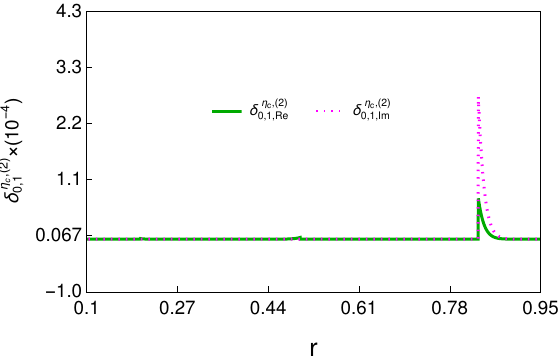}
\hspace{0cm}\includegraphics[width=0.495\textwidth]{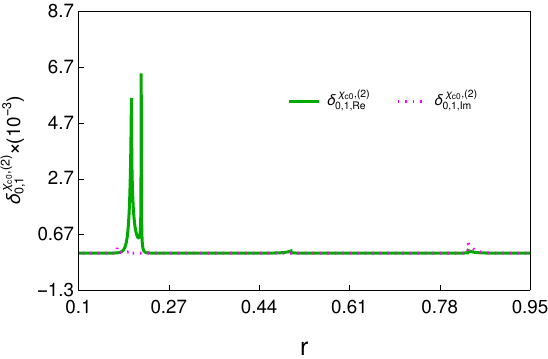}
\hspace{0cm}\includegraphics[width=0.495\textwidth]{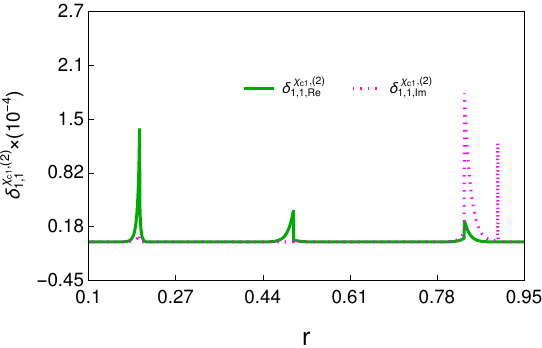}
\hspace{0cm}\includegraphics[width=0.495\textwidth]{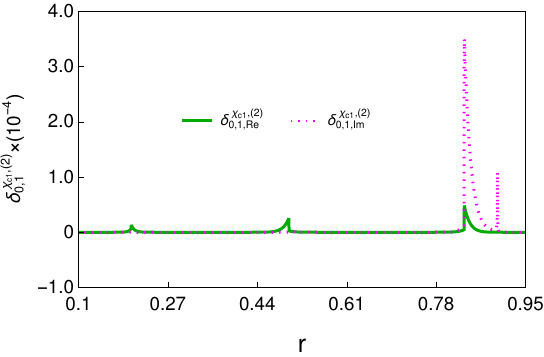}
\hspace{0cm}\includegraphics[width=0.495\textwidth]{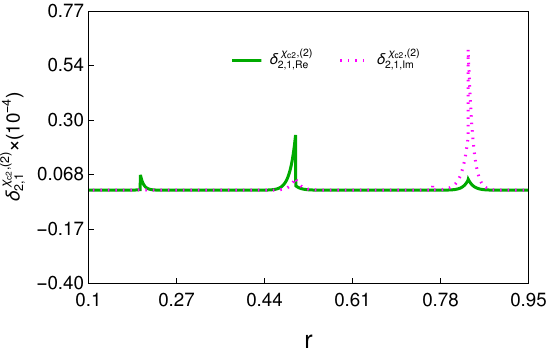}
\hspace{0cm}\includegraphics[width=0.495\textwidth]{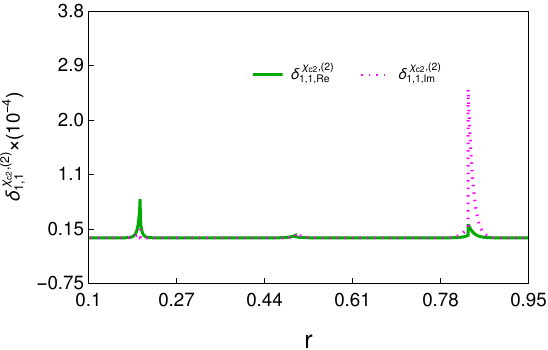}
\hspace{0cm}\includegraphics[width=0.495\textwidth]{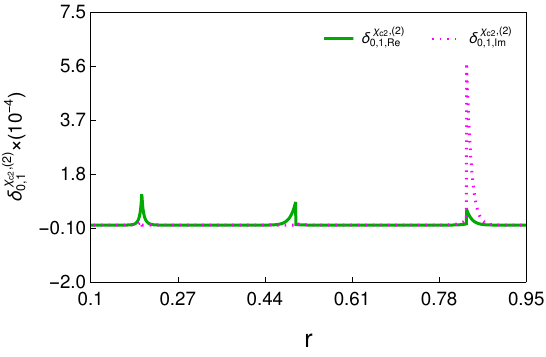}
\caption{\label{fig:loop2:abs}
The relative errors between the combined asymptotic expansions and the exact results of $c^{H,(n)}_{\lambda_1,\lambda_2}(r)$. $\delta^{H}_{\lambda_1,\lambda_2,\rm Re}$ and $\delta^{H}_{\lambda_1,\lambda_2,\rm Im}$ are defined in Eq.~(\ref{eq:loop2:rel:err}).}
\end{center}
\end{figure*}

We now compare the combined asymptotic expressions with the exact results for $c^{H,(2)}_{\lambda_1,\lambda_2}$. Specifically, we use the expansions at $r_0=0$ for $0 < r\leq 1/5$, the expansions at $r_0=1/3$ for $1/5 < r \leq 1/2$, the expansions at $r_0=2/3$ for $1/2 < r \leq 5/6$, and the expansion at $r_0=1$ for $5/6 < r < 1$. The comparisons are illustrated in Fig.~\ref{fig:loop2:abs}, where the relative errors are defined as
\begin{eqnarray}~\label{eq:loop2:rel:err}
\delta^{H}_{\lambda_1,\lambda_2,\rm Re} &=& \frac{{\rm Re}[g^{H,(2)}_{\lambda_1,\lambda_2}(15,2)]-{\rm Re}[g^{H,(n)}_{\lambda_1,\lambda_2;\rm tot}]}{{\rm Re}[g^{H,(n)}_{\lambda_1,\lambda_2;\rm tot}]},\nonumber\\
\delta^{H}_{\lambda_1,\lambda_2,\rm Im} &=& \frac{{\rm Im}[g^{H,(2)}_{\lambda_1,\lambda_2}(15,2)]-{\rm Im}[g^{H,(n)}_{\lambda_1,\lambda_2;\rm tot}]}{{\rm Im}[g^{H,(n)}_{\lambda_1,\lambda_2;\rm tot}]}.
\end{eqnarray}

From the figure, we observe that the combined asymptotic expression approximates the exact results accurately over the entire range $0\leq r\leq 1$. The relative error remains below $0.1\%$ over most of the range, rising to less than $1\%$ only within a very limited interval of $r$ values.  

\section{Phenomenology~\label{sec:phe}}

\subsection{Input parameters}

Before presenting phenomenological predictions for $e^+e^-\to H+\gamma$ at B factories, we specify the input parameters used in our analysis. The charm quark mass is taken as $m_c = 1.5 \pm 0.2$ GeV. The QED coupling constant is set to $\alpha(\sqrt{s} = 10.6 {\rm GeV}) = 1/130.9$. 

We take $n_f=4$ active quark flavors, and determine the strong coupling constant $\alpha_s$ at various renormalization scales $\mu_R$ by solving the renormalization group equation using \texttt{RunDec}~\cite{Chetyrkin:2000yt}. Uncertainties in the cross sections and the angular distribution parameters $\alpha^H_\theta$ are estimated by varying $\mu_R$ from $2m_c=3$ GeV to $\sqrt{s}$, with the central value  fixed at $\mu_R=\sqrt{s}/2$. 

In this work, the factorization scale is fixed at $\mu_\Lambda=1$ GeV. The corresponding LDMEs at this scale are taken from Refs.~\cite{Chung:2008km,Bodwin:2007fz,Bodwin:2006dn}:
\begin{eqnarray}
&&|\langle{\mathcal{O}_{\eta_c}}\rangle(1\,{\rm GeV})|^2=0.437~\mathrm{GeV}^3,\nonumber\\
&&|\langle{\mathcal{O}_{\chi_{c0}}} \rangle(1\,{\rm GeV})|^2=0.051~\mathrm{GeV}^5,\nonumber\\
&&|\langle{\mathcal{O}_{\chi_{c1}}} \rangle(1\,{\rm GeV})|^2=0.060~\mathrm{GeV}^5,\nonumber\\
&&|\langle{\mathcal{O}_{\chi_{c2}}} \rangle(1\,{\rm GeV})|^2=0.068~\mathrm{GeV}^5.
\end{eqnarray}

\subsection{Prediction for the cross sections}

We now assemble all components to predict the cross sections for $e^+e^-\to H+\gamma$, as summarized in Table~\ref{tab:cross:section}. The corresponding experimental measurements from the {\tt Belle} collaboration~\cite{Belle:2018jqa} are also included for comparison. The uncertainty in the LO prediction arises solely from the charm quark mass, while the NLO and NNLO theoretical uncertainties receive contributions from both the charm quark mass and variations of the renormalization scale.


\begin{table}[!h]
\begin{center}
\caption{Cross sections (in units of fb) of $e^+e^-\to H+\gamma$ at $\sqrt{s} = 10.58$ GeV. The first uncertainties originate from the charm quark mass $m_c = 1.5 \pm 0.2$ GeV, and the second from varying the renormalization scale $\mu_R$ between $3$ GeV to $\sqrt{s}$, with the the central value set to $\sqrt{s}/2$.}
\label{tab:cross:section}
\begin{tabular}{|c|c|c|c|c|c|}
\hline
 & LO & NLO & NNLO & {\tt Belle}~\cite{Belle:2018jqa}\\ \hline
$\eta_{c}+\gamma$ & $77.19^{+13.81}_{-10.78}$ & $63.98^{+13.51+2.29}_{-10.24-2.80}$ & $39.17^{+4.24+8.51}_{-3.73-11.66}$& $<21.1$ at $90\%$ C.L.\\ 
$ \chi_{c0}+\gamma$ & $0.91^{+0.68}_{-0.38}$ & $1.04^{+0.73+0.03}_{-0.41-0.02}$ & $1.08^{+0.73+0.01}_{-0.41-0.02}$&$<205.9$ at $90\%$ C.L. \\ 
$\chi_{c1}+\gamma$ & $12.04^{+5.72}_{-3.38}$ & $9.36^{+4.91+0.46}_{-2.82-0.57}$ & $7.22^{+3.53+0.80}_{-2.08-0.98}$&$17.3^{+4.2}_{-3.9}\pm1.7$ \\ 
$\chi_{c2}+\gamma$ & $5.39^{+2.23}_{-1.31}$ & $2.16^{+0.86+0.56}_{-0.50-0.68}$ & $0.58^{+0.13+0.65}_{-0.09-0.71}$&$<5.7$ at $90\%$ C.L. \\ \hline
\end{tabular}
\end{center}
\end{table}

Several observations can be drawn from the table. First, the $\mathcal{O}(\alpha_s)$ correction to $\eta_c+\gamma$ production is negative and moderate, whereas the $\mathcal{O}(\alpha_s^2)$ correction is sizable, reducing the cross section by about half relative to the LO prediction.  For $\chi_{c2}+\gamma$, both the $\mathcal{O}(\alpha_s)$ and $\mathcal{O}(\alpha_s^2)$ corrections are negative and substantial, making the NNLO prediction an order of magnitude smaller than the LO result. In contrast, the perturbative corrections are moderate for $\chi_{c1}+\gamma$ and small for $\chi_{c0}+\gamma$ indicating a well-behaved perturbative expansion in these channels.~\footnote{Using the same set of parameters, our $\mathcal{O}(\alpha_s^2)$ correction for $\eta_c+\gamma$ agrees well with the results presented in Refs.~\cite{Chen:2017pyi,Yu:2020tri}. For $\chi_{c0}+\gamma$ and $\chi_{c2}+\gamma$, our $\mathcal{O}(\alpha_s^2)$ corrections show slight differences compared to those given in Ref.~\cite{Sang:2020fql}, while a more pronounced discrepancy is observed for the $\chi_{c1}+\gamma$ channel. The deviations likely originate from an underestimation of numerical precision in Ref.~\cite{Sang:2020fql}.}

Second, the theoretical predictions exhibit large uncertainties due to the charm quark mass, particularly for $\chi_{cJ}$ production. This reflects the strong sensitivity of the cross sections to $m_c$, scaling as $\sigma[\eta_c+\gamma]\propto 1/m_c$ and $\sigma[\chi_{cJ}+\gamma]\propto 1/m_c^3$. The uncertainties from $\mu_R$ variations are small for $\chi_{c1}$, moderate for $\chi_{c0}$, and significant for $\eta_c$ and $\chi_{c2}$, consistent with the relative size of the QCD corrections in each channel. 

Third, the cross section for $\eta_c$ production is significantly larger than for $\chi_{cJ}$ production, primarily due to the larger LDME for $\eta_c$. Moreover, the cross section for $\chi_{c1}+\gamma$ is roughly an order of magnitude larger than for $\chi_{c0}$ and $\chi_{c2}$. This could partly explain why only the $\chi_{c1}+\gamma$ channel has been observed experimentally so far, while the other two channels remain undetected.

A direct comparison with {\tt Belle} data is straightforward. The {\tt Belle} collaboration reports $\sigma[\chi_{c1}+\gamma] = 17.3^{+4.2}_{-3.9}({\rm stat.})\pm 1.7({\rm syst.})$ fb. While the central value of the theoretical prediction is approximately half of the experimental value, the two are consistent within $2\sigma$ when both theoretical and experimental uncertainties are taken into account.~\footnote{We note that adopting the LDMEs from Ref.~\cite{Sang:2020fql}, based on the Buchm\"uller-Tye potential model, which gives a value roughly $1.8$ times larger for $|\langle \mathcal{O}_{\chi_{c1}}\rangle|^2$ than used here, would reduce the discrepancy in central values between theory and experiment.} 

\begin{figure*}[!h]
\begin{center}
\hspace{0cm}\includegraphics[width=0.495\textwidth]{etacpolemass.pdf}
\hspace{0cm}\includegraphics[width=0.495\textwidth]{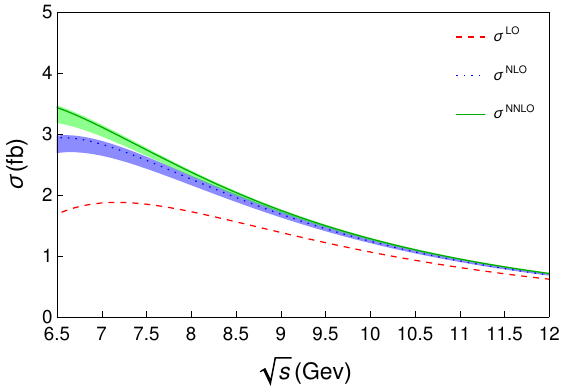}
\hspace{0cm}\includegraphics[width=0.495\textwidth]{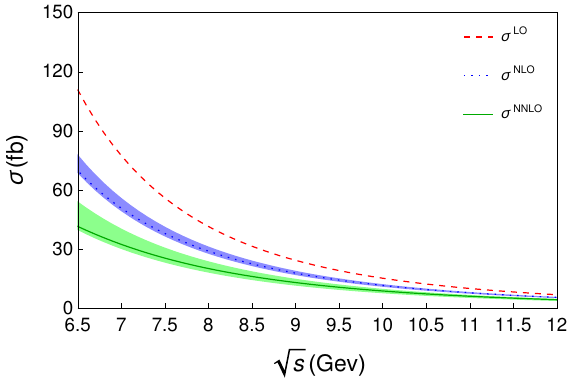}
\hspace{0cm}\includegraphics[width=0.495\textwidth]{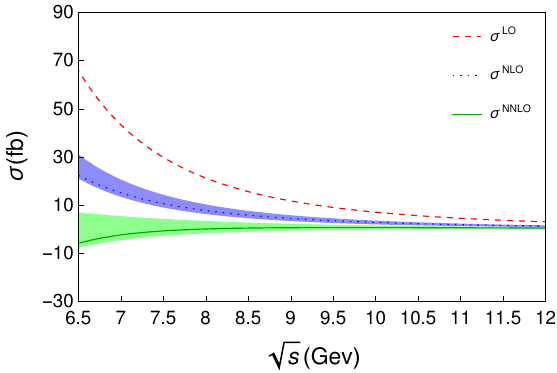}
\caption{\label{fig:cross:section}
Cross sections for $e^+e^-\to H+\gamma$ as functions of $\sqrt{s}$ at different perturbative orders, computed with $m_c=1.5$ GeV. The shaded band indicates renormalization scale uncertainties from varying $\mu_R$ from $3$ GeV to $\sqrt{s}$, with central values $\mu_R=\sqrt{s}/2$.}
\end{center}
\end{figure*}

In Fig.~\ref{fig:cross:section}, we show the cross sections as functions of  $\sqrt{s}$ at different perturbative orders. Here, $m_c$ is fixed at $1.5$ GeV, and uncertainties from the renormalization scale $\mu_R$ are included.

For $\eta_c$ and $\chi_{c1}$ production, the cross sections decrease rapidly with increasing $\sqrt{s}$ at all considered order. In the case of $\chi_{c0}+\gamma$, the LO and NLO predictions initially rise and then fall with $\sqrt{s}$, whereas the NNLO result decreases monotonically for $\sqrt{s}>6.5$ GeV. For $\chi_{c2}+\gamma$, although the LO and NLO cross sections fall with $\sqrt{s}$, the NNLO prediction remains small in magnitude and exhibits an anomalous increase with $\sqrt{s}$. More notably, the NNLO prediction becomes negative at lower $\sqrt{s}$.  All these features point to poor perturbative convergence in the $\chi_{c2}+\gamma$ channel.

\subsection{Prediction for the angular distribution parameters $\alpha^H_\theta$}

In this subsection, we present theoretical predictions for the angular distribution parameters $\alpha^H_\theta$ defined in Eq.~\eqref{eq:ang:dis}. To obtain predictions at different perturbative accuracies, we expand $\alpha^H_\theta$ as a series in $\alpha_s$. 
The corresponding results are summarized in Table~\ref{tab:ang:dis}, where the two uncertainties are affiliated with variations in the charm quark mass and the renormalization scale.

Owing to the unique helicity structure of the $\eta_c+\gamma$ and $\chi_{c0}+\gamma$, the parameters $\alpha^{\eta_c}_\theta$ and $\alpha^{\chi_{c0}}_\theta$ are identically equal to 1 at all perturbative orders. In contrast, the parameters $\alpha^{\chi_{c1}}_\theta$ and $\alpha^{\chi_{c2}}_\theta$ receive non-trivial corrections from higher-order QCD corrections.

Unlike the cross-section predictions, the $\mathcal{O}(\alpha_s)$ and $\mathcal{O}(\alpha_s^2)$ corrections to $\alpha^{\chi_{c1}}_\theta$ and $\alpha^{\chi_{c2}}_\theta$ are very small, reflecting good perturbative convergence. Furthermore, the uncertainties in these parameters induced by renormalization scale variation are minimal, while those arising from the charm quark mass remain moderate. This behavior can be attributed to a substantial cancellation of $\mu_R$ dependence between the numerator and denominator in the definition of $\alpha^H_\theta$. A similar cancellation occurs for the charm mass.

More importantly, the theoretical predictions for $\alpha^{\chi_{c1}}_\theta$ and $\alpha^{\chi_{c2}}_\theta$ are insensitive to the NRQCD LDMEs, enabling precise predictions and offering an ideal setting to test the NRQCD factorization approach. With the high luminosity expected at {\tt Belle 2}, we anticipate that future experimental measurements of these parameters will provide a clear test of NRQCD factorization.

\begin{table}[!h]
\begin{center}
\caption{Theoretical predictions for the angular distribution parameters $\alpha^H_\theta$. Parameter choices and uncertainty sources are the same as in Table~\ref{tab:cross:section}.}
\label{tab:ang:dis}
\setlength{\tabcolsep}{2pt}
\begin{tabular}{|c|c|c|c|c|c|c|c|c|}
\hline
 & {LO} &{NLO}& {NLO} \\ \hline
$\eta_{c}+\gamma$ &$1$ &$1$ &$1$ \\ 
$ \chi_{c0}+\gamma$ &$1$ &$1$ &$1$ \\ 
$\chi_{c1}+\gamma$ &  $0.72^{+0.06+0.00}_{-0.07-0.00}$  &   $0.71^{+0.06+0.01}_{-0.07-0.01}$  & $0.71^{+0.06+0.01}_{-0.07-0.01}$ \\ 
$\chi_{c2}+\gamma$ & $0.37^{+0.11+0.00}_{-0.10-0.00}$ &  $0.41^{+0.12+0.01}_{-0.11-0.01}$ &   $0.47^{+0.12+0.03}_{-0.11-0.02}$\\  \hline
\end{tabular}
\end{center}
\end{table}

\section{summary~\label{sec:sum}}

In this work, we investigate the processes $e^+e^-\to \eta_c/\chi_{cJ}+\gamma$ at B factories, computing the helicity amplitudes through $\mathcal{O}(\alpha_s^2)$ within the NRQCD factorization framework. 

We compute the SDCs up to $\mathcal{O}(\alpha_s^2)$ as asymptotic expansions around $r=0, 1/3, 2/3, 1$, using the method of differential equations. At $\mathcal{O}(\alpha_s)$ the expansions around $r=0$ approximate the exact results well over most of the $r$ range, with notable deviations only appearing in the imaginary part as $r\to 1$. At $\mathcal{O}(\alpha_s^2)$, however, the expansions around $r=0$ are accurate only for $r<0.27$ in the $\eta_c+\gamma$ channel, and for even smaller $r$ in the other channels. Beyond these limits the deviations increase rapidly—particularly for the $\chi_{c0}+\gamma$ channel, whose exact result exhibits singular behaviour near $r=1/3$. 
By combining asymptotic expansions around all four points, we construct composite asymptotic expressions that reproduce the exact results accurately over the whole interval $0\leq r\leq1$. The relative error stays below $0.1\%$ over most of the domain and remains under $1\%$ elsewhere.

Using these asymptotic expressions, we compute the unpolarized cross sections. The QCD corrections are found to be small for $\chi_{c0}+\gamma$, moderate for $\chi_{c1}+\gamma$, and substantial for $\eta_c+\gamma$ and $\chi_{c2}+\gamma$. In particular, both the $\mathcal{O}(\alpha_s)$ and $\mathcal{O}(\alpha_s^2)$ corrections to  $\chi_{c2}+\gamma$ are negative and large, reducing the NNLO prediction an order of magnitude smaller than the LO result. Comparison with {\tt Belle} measurement for $\chi_{c1}+\gamma$ shows consistency between theory and experiment within $2\sigma$, showing agreement between theory and experimen.

We also predict the angular distribution parameters $\alpha^H_\theta$, which are insensitive to the NRQCD LDMEs and exhibit small theoretical uncertainties.
With the high luminosity expected at {\tt Belle 2}, we anticipate that future experimental measurements of these parameters will provide a clear test
of NRQCD factorization.

\section{Acknowledgments} 
The work of C. Li. and W.-L. S. is supported by the
National Natural Science Foundation of China under Grants
No. 12375079, and the Natural Science
Foundation of ChongQing under Grant No. CSTB2023
NSCQ-MSX0132. 
The work of H.-F. Zhang. is supported by the
National Natural Science Foundation of China under Grants
No. 12565012.


\appendix
\section{Asymptotic expressions at $r=0$}

\subsection{Asymptotic expressions for NLO SDC up to $\mathcal{O}(r^1)$~\label{app:1loop}}

\begin{eqnarray}\label{one-ex1}
f^{\eta_c,(1)}_{0,1}(0,0)&=&\left(-\frac{9}{4}-\frac{\pi ^2}{12}-\frac{3 \ln^2 2}{4}+\frac{9 \ln 2}{4}\right) C_F+i \left(\frac{1}{2} \pi  \ln 2-\frac{3 \pi }{4}\right)C_F,\nonumber \\
f^{\chi_{c0},(1)}_{0,1}(0,0)&=&\left(-\frac{\pi^2}{12}-\frac{3 \ln^2 2}{4}+\frac{11 \ln 2}{4}\right) C_F+i \left(\frac{1}{2} \pi  \ln 2-\frac{\pi }{4}\right)C_F,\nonumber \\
f^{\chi_{c1},(1)}_{1,1}(0,0)&=&\left(-\frac{3}{2}-\frac{\pi ^2}{12}-\frac{3 \ln^2 2}{4}+\frac{5 \ln 2}{4}\right) C_F+i \left(\frac{1}{2} \pi  \ln 2-\frac{3 \pi }{4}\right)C_F,\nonumber \\
f^{\chi_{c1},(1)}_{0,1}(0,0)&=&\left(-\frac{7}{4}-\frac{\pi ^2}{12}-\frac{3 \ln^2 2}{4}+\frac{5 \ln 2}{4}\right) C_F+i \left(\frac{1}{2} \pi  \ln 2-\frac{3 \pi }{4}\right)C_F,\nonumber \\
f^{\chi_{c2},(1)}_{2,1}(0,0)&=&\left(-2-\frac{\pi^2}{3}-3 \ln ^2 2+6\ln 2\right) C_F+i (2 \pi  \ln 2-\pi )C_F,\nonumber \\
f^{\chi_{c2},(1)}_{1,1}(0,0)&=&\left(-\frac{\pi^2}{12}-\frac{3 \ln^2 2}{4}-\frac{7 \ln 2}{4}\right) C_F+i \left(\frac{\pi}{4}+\frac{1}{2} \pi  \ln 2\right)C_F,\nonumber \\
f^{\chi_{c2},(1)}_{0,1}(0,0)&=&\left(-\frac{3}{2}-\frac{\pi ^2}{12}-\frac{3 \ln^2 2}{4}-\frac{\ln 2}{4}\right) C_F+i \left(\frac{1}{2} \pi  \ln 2-\frac{\pi }{4}\right)C_F,
\end{eqnarray}

\begin{eqnarray}\label{one-ex1}
f^{\chi_{c1},(1)}_{1,1}(1,2)&=&f^{\chi_{c2},(1)}_{2,1}(1,2)=f^{\chi_{c2},(1)}_{1,1}(1,2)=0,\nonumber \\
f^{\eta_c,(1)}_{0,1}(1,2)&=&-\frac{C_F}{8},f^{\chi_{c0},(1)}_{0,1}(1,2)=-\frac{3C_F}{8},f^{\chi_{c1},(1)}_{0,1}(1,2)=-\frac{C_F}{8},f^{\chi_{c2},(1)}_{0,1}(1,2)=-\frac{3 C_F}{8},\nonumber\\
\end{eqnarray}

\begin{eqnarray}\label{one-ex1}
f^{\eta_c,(1)}_{0,1}(1,1)&=&\left(\ln 2-\frac{1}{2}\right)C_F-\frac{i}{4} \pi   C_F\nonumber \\
f^{\chi_{c0},(1)}_{0,1}(1,1)&=&\left(\frac{5 \ln 2}{2}-\frac{3}{2}\right)C_F-\frac{3i}{4} \pi   C_F,\nonumber \\
f^{\chi_{c1},(1)}_{1,1}(1,1)&=&\left(\frac{5}{8}+\frac{\ln 2}{4}\right) C_F,\nonumber \\
f^{\chi_{c1},(1)}_{0,1}(1,1)&=&\ln 2 C_F-\frac{i}{4} \pi  C_F,\nonumber \\
f^{\chi_{c2},(1)}_{2,1}(1,1)&=&\left(4\ln 2-\frac{5}{2}\right) C_F,\nonumber \\
f^{\chi_{c2},(1)}_{1,1}(1,1)&=&\left(\frac{23}{8}-\frac{17\ln 2}{4}\right) C_F,\nonumber \\
f^{\chi_{c2},(1)}_{0,1}(1,1)&=&\left(\frac{5 \ln 2}{2}-\frac{3}{2}\right)C_F-\frac{3i}{4} \pi   C_F,
\end{eqnarray}

\begin{eqnarray}\label{one-ex1}
f^{\eta_c,(1)}_{0,1}(1,0)&=&\left(\frac{3}{8}-\frac{\pi ^2}{12}-\frac{5 \ln^2 2}{4}+\frac{3 \ln 2}{2}\right) C_F+i \left(\pi  \ln 2-\frac{\pi }{2}\right)C_F,\nonumber \\
f^{\chi_{c0},(1)}_{0,1}(1,0)&=&\left(\frac{17}{4}-\frac{\pi ^2}{6}-3 \ln ^2 2+7\ln 2\right) C_F+i \left(\frac{5}{2} \pi  \ln 2-\frac{3 \pi }{2}\right)C_F,\nonumber \\
f^{\chi_{c1},(1)}_{1,1}(1,0)&=&\left(-\frac{1}{4}-\frac{\pi ^2}{24}-\frac{3 \ln^2 2}{8}-\frac{\ln 2}{8}\right) C_F+i \left(\frac{5 \pi}{8}+\frac{1}{4} \pi  \ln 2\right)C_F,\nonumber \\
f^{\chi_{c1},(1)}_{0,1}(1,0)&=&\left(-\frac{1}{8}-\frac{\pi ^2}{12}-\frac{5 \ln^2 2}{4}+2 \ln 2\right)C_F+i\pi   \ln 2 C_F,\nonumber \\
f^{\chi_{c2},(1)}_{2,1}(1,0)&=&\left(\frac{5}{2}-\frac{2 \pi ^2}{3}-6 \ln^2 2+10 \ln 2\right)C_F+i \left(4 \pi  \ln 2-\frac{5 \pi }{2}\right)C_F,\nonumber \\
f^{\chi_{c2},(1)}_{1,1}(1,0)&=&\left(\frac{17 \pi^2}{24}+\frac{51 \ln^2 2}{8}-\frac{115 \ln 2}{8}\right) C_F+i \left(\frac{23 \pi}{8}-\frac{17}{4} \pi  \ln 2\right)C_F,\nonumber \\
f^{\chi_{c2},(1)}_{0,1}(1,0)&=&\left(-\frac{5}{2}-\frac{\pi ^2}{6}-3 \ln^2 2+10 \ln 2\right)C_F+i \left(\frac{5}{2} \pi  \ln 2-\frac{3 \pi }{2}\right)C_F.
\end{eqnarray}

\subsection{Asymptotic expressions for NNLO SDC up to $\mathcal{O}(r^1)$~\label{app:2loop}}

\begin{eqnarray}\label{one-ex1}
f^{\eta_c,(2)}_{0,1}(0,1)&=& \left(\frac{3 \zeta(3)}{32}-\frac{17}{96}+\frac{11 \pi ^2}{288}+\frac{11\ln ^2 2}{48}-\frac{17\ln 2}{72}\right)C_A C_F\nonumber\\
&+&\left(-\frac{7 \zeta(3)}{32}+\frac{51}{32}+\frac{5 \pi ^2}{48}-\frac{\ln^3 2}{6}+\frac{13 \ln^2 2}{8}-\frac{39 \ln 2}{16}-\frac{\pi^2 \ln 2}{16} \right)C_F^2\nonumber\\
&+&\left(\frac{1}{6}-\frac{\pi^2}{18}-\frac{\ln^2 2}{3}+\frac{2 \ln 2}{9}\right) C_F T_f\nonumber\\
&+&\left[\left(\frac{55 \zeta(3)}{8}-\frac{35 \pi^2}{48}+\frac{5 \ln^3 2}{6}-\frac{15 \ln^2 2}{4}+\frac{5\ln 2}{2}+\frac{5\pi ^2\ln2}{24} \right) C_F T_f\right]_{\rm lbl}\nonumber\\
&+&i\left\{\left[\left(-\frac{5 \pi^3}{24}-\frac{5\pi \ln ^2 2}{4} +\frac{15\pi\ln 2}{4} \right) C_FT_f\right]_{\rm lbl}+\left(\frac{9 \pi}{16}-\frac{\pi^3}{48}+\frac{\pi \ln ^2 2}{8} -\pi  \ln 2\right) C_F^2\right\},\nonumber \\
f^{\chi_{c0},(2)}_{0,1}(0,1)&=& \left(\frac{3 \zeta(3)}{32}-\frac{383}{288}+\frac{11 \pi^2}{288}+\frac{11 \ln^2 2}{48}-\frac{10 \ln 2}{9}\right)C_A C_F\nonumber\\
&+&\left(-\frac{7 \zeta(3)}{32}-\frac{11}{32}+\frac{5 \pi ^2}{48}-\frac{\ln^3 2}{6}+\frac{\ln^2 2}{8}+\frac{5 \ln 2}{16}-\frac{1}{16} \pi^2 \ln 2\right)C_F^2\nonumber\\
&+&\left(\frac{31}{18}-\frac{\pi^2}{18}-\frac{\log^2(2)}{3}+\frac{20 \ln2}{9}\right) C_F T_f\nonumber\\
&+&\left[\left(\frac{55 \zeta(3)}{8}-\frac{25 \pi^2}{48}+\frac{5 \ln^3 2}{6}-\frac{5 \ln^2 2}{2}-\frac{15 \ln2}{2}+\frac{5}{24} \pi ^2\ln2\right) C_F T_f\right]_{\rm lbl}\nonumber\\
&+&i\left\{\left[\left(\frac{5 \pi}{4}-\frac{5 \pi^3}{24}-\frac{5}{4} \pi \ln ^2 2+\frac{15}{4} \pi\ln 2\right) C_FT_f\right]_{\rm lbl}+\left(-\frac{7 \pi}{16}-\frac{\pi^3}{48}+\frac{1}{8} \pi \ln ^2 2\right.\right.\nonumber\\
&+&\left.\left.\frac{1}{4} \pi \ln 2\right)C_F^2\right\},\nonumber \\
f^{\chi_{c1},(2)}_{1,1}(0,1)&=& \left(\frac{3 \zeta(3)}{32}-\frac{73}{96}-\frac{19 \pi^2}{288}-\frac{\ln^2 2}{48}+\frac{139\ln 2}{72}\right)C_AC_F\nonumber\\
&+&\left(-\frac{7 \zeta(3)}{32}+\frac{31}{32}+\frac{9 \pi ^2}{32}-\frac{\ln^3 2}{6}+\frac{7 \ln^2 2}{8}-\frac{51 \ln 2}{16}-\frac{1}{16} \pi^2 \ln 2\right)C_F^2\nonumber\\
&+&\left(\frac{5}{6}-\frac{\pi^2}{18}-\frac{\ln^2 2}{3}-\frac{10 \ln 2}{9}\right)C_F T_f\nonumber\\
&+&\left[\left(-\frac{55 \zeta(3)}{4}+\frac{5 \pi^2}{2}-\frac{5 \ln^3 2}{3}+\frac{55 \ln^2 2}{4}-\frac{25 \ln2}{2}-\frac{5}{12} \pi ^2\ln2\right)C_F T_f\right]_{\rm lbl}\nonumber\\
&+&i \left\{\left[\left(\frac{1}{4} \pi \ln 2-\frac{\pi}{8}\right) C_AC_F+\left(\frac{5 \pi^3}{12}+\frac{5}{2} \pi \ln ^2 2-\frac{15}{2} \pi\ln 2\right) C_FT_f\right]_{\rm lbl}\right.\nonumber\\
&+&\left.\left(\frac{9 \pi}{16}-\frac{\pi^3}{48}+\frac{1}{8} \pi \ln ^2 2-\pi  \ln 2\right) C_F^2\right\},\nonumber 
\end{eqnarray}

\begin{eqnarray}\label{one-ex1}
f^{\chi_{c1},(2)}_{0,1}(0,1)&=& \left(\frac{3 \zeta(3)}{32}-\frac{61}{96}+\frac{11 \pi ^2}{288}+\frac{11\ln ^2 2}{48}+\frac{49\ln 2}{72}\right)C_A C_F\nonumber\\
&+&\left(-\frac{7 \zeta(3)}{32}+\frac{39}{32}+\frac{7 \pi ^2}{96}-\frac{\ln^3 2}{6}+\frac{11 \ln^2 2}{8}-\frac{35 \ln 2}{16}-\frac{1}{16} \pi^2 \ln 2\right)C_F^2\nonumber\\
&+&\left(\frac{5}{6}-\frac{\pi^2}{18}-\frac{\ln^2 2}{3}-\frac{10 \ln2}{9}\right)C_FT_f \nonumber\\
&+&\left[\left(\frac{55 \zeta(3)}{8}-\frac{15 \pi^2}{16}+\frac{5 \ln^3 2}{6}-\frac{25 \ln^2 2}{4}+\frac{15 \ln2}{2}+\frac{5}{24} \pi ^2\ln2\right)C_FT_f\right] _{\rm lbl}\nonumber\\
&+&i\left\{\left[\left(-\frac{5 \pi^3}{24}-\frac{5}{4} \pi \ln ^2 2+\frac{15}{4} \pi\ln 2\right) C_FT_f\right]_{\rm lbl}+\left(\frac{9 \pi}{16}-\frac{\pi^3}{48}+\frac{1}{8} \pi \ln ^2 2-\pi  \ln 2\right) C_F^2\right\},\nonumber\\
f^{\chi_{c2},(2)}_{2,1}(0,1)&=&\left(\frac{3 \zeta(3)}{8}-\frac{13}{9}+\frac{5 \pi ^2}{72}+\frac{11 \ln^2 2}{12}+\frac{259 \ln 2}{72}\right)C_A C_F\nonumber\\
&+&\left(-\frac{37 \zeta(3)}{8}+2+\frac{11 \pi^2}{16}-\frac{5 \ln^3 2}{3}+\frac{31 \ln^2 2}{8}-\frac{61 \ln 2}{8}-\frac{1}{8} \pi ^2\ln 2\right)C_F^2\nonumber\\
&+&\left(\frac{8}{9}-\frac{2 \pi^2}{9}-\frac{4 \log^2(2)}{3}-\frac{22 \ln2}{9}\right) C_F T_f\nonumber\\
&+&\left[\left(\frac{55 \zeta(3)}{8}-\frac{19}{36}-\frac{11 \pi ^2}{24}+\frac{5\ln ^3 2}{6}-\frac{91\ln ^2 2}{12}+\frac{7\ln2}{9}+\frac{5}{24}\pi ^2 \ln2\right)C_F T_f\right]_{\rm lbl}\nonumber\\
&+&i \left\{\left(\frac{1}{2} \pi \ln 2-\frac{\pi}{4}\right) C_AC_F+\left[\left(\frac{31 \pi}{9}-\frac{5 \pi^3}{24}-\frac{5}{4} \pi \ln ^2 2+\frac{1}{2} \pi \ln 2\right) C_FT_f\right]_{\rm lbl}\right.\nonumber\\
&+&\left.\left(\frac{3 \pi}{4}+\frac{\pi^3}{24}+\frac{5}{4} \pi \ln ^2 2-\frac{11}{4} \pi\ln 2\right)C_F^2\right\},\nonumber \\
f^{\chi_{c2},(2)}_{1,1}(0,1)&=&\left(\frac{3 \zeta(3)}{32}+\frac{533}{288}+\frac{5 \pi ^2}{288}+\frac{11\ln ^2 2}{48}+\frac{61\ln 2}{72}\right)C_A C_F\nonumber\\
&+&\left(-\frac{7 \zeta(3)}{32}-\frac{49}{32}-\frac{11 \pi ^2}{96}-\frac{\ln^3 2}{6}-\frac{5 \ln^2 2}{8}+\frac{41 \ln 2}{16}-\frac{1}{16} \pi^2 \ln 2\right)C_F^2\nonumber\\
&+&\left(-\frac{25}{18}-\frac{\pi^2}{18}-\frac{\ln^2 2}{3}-\frac{22 \ln2}{9}\right)C_FT_f \nonumber\\
&+&\left[\left(-\frac{55 \zeta(3)}{4}-\frac{35}{9}+\frac{35 \pi ^2}{24}-\frac{5 \ln^3 2}{3}+\frac{145 \ln^2 2}{12}+\frac{35 \ln2}{9}-\frac{5}{12} \pi ^2\ln2\right)C_FT_f \right]_{\rm lbl}\nonumber\\
&+&i \left\{\left(\frac{\pi}{8}-\frac{1}{4} \pi  \ln 2\right) C_AC_F+\left[\left(-\frac{95 \pi}{18}+\frac{5 \pi^3}{12}+\frac{5}{2} \pi \ln ^2 2\right) C_FT_f\right]_{\rm lbl}\right.\nonumber\\
&+&\left.\left(\frac{\pi}{16}-\frac{\pi^3}{48}+\frac{1}{8} \pi \ln ^2 2+\frac{1}{2} \pi \ln 2\right)C_F^2\right\}, 
\end{eqnarray}

\begin{eqnarray}\label{one-ex1}
f^{\chi_{c2},(2)}_{0,1}(0,1)&=& \left(\frac{3 \zeta(3)}{32}+\frac{13}{288}+\frac{11 \pi ^2}{288}+\frac{11\ln ^2 2}{48}+\frac{59\ln 2}{36}\right)C_A C_F\nonumber\\
&+&\left(-\frac{7 \zeta(3)}{32}+\frac{1}{32}-\frac{17 \pi ^2}{96}-\frac{\ln^3 2}{6}-\frac{5 \ln^2 2}{8}+\frac{41 \ln 2}{16}-\frac{1}{16} \pi^2 \ln 2\right)C_F^2\nonumber\\
&+&\left(-\frac{5}{18}-\frac{\pi^2}{18}-\frac{\ln^2 2}{3}-\frac{16 \ln2}{9}\right)C_F T_f\nonumber\\
&+&\left[\left(\frac{55 \zeta(3)}{8}+\frac{10}{3}-\frac{55 \pi ^2}{48}+\frac{5 \ln^3 2}{6}-\frac{15 \ln^2 2}{2}+\frac{25 \ln2}{6}+\frac{5}{24} \pi ^2\ln2\right)C_F T_f\right]_{\rm lbl}\nonumber\\
&+&i\left\{\left[\left(\frac{5 \pi}{12}-\frac{5 \pi^3}{24}-\frac{5}{4} \pi \ln ^2 2+\frac{15}{4} \pi\ln 2\right) C_FT_f\right]_{\rm lbl}+\left(-\frac{7 \pi}{16}-\frac{\pi^3}{48}+\frac{1}{8} \pi \ln ^2 2\right.\right.\nonumber\\
&+&\left.\left.\frac{1}{4} \pi \ln 2\right)C_F^2\right\},
\end{eqnarray}

\begin{eqnarray}\label{one-ex1}
f^{\eta_c,(2)}_{0,1}(0,0)&=&-5.2850920 C_AC_F+(3.8227005C_F T_f)_{\rm lbl}-1.7158929 C_FT_f-13.433618 C_F^2\nonumber \\&+&i \left[-4.58077471 C_AC_F+(7.4932279C_F T_f)_{\rm lbl}+6.6283558 C_FT_f+3.0379251C_F^2\right],\nonumber \\
f^{\chi_{c0},(2)}_{0,1}(0,0)&=&-2.6321553 C_A C_F-(5.3158292C_F T_f)_{\rm lbl}+1.0068682 C_FT_f-1.7206212 C_F^2\nonumber \\
&+&i \left[-2.6301256 C_AC_F-(0.6149258C_F T_f)_{\rm lbl}+4.9939479 C_FT_f+0.6753696C_F^2\right],\nonumber \\
f^{\chi_{c1},(2)}_{1,1}(0,0)&=&-3.6278248 C_AC_F+(2.9172125C_F T_f)_{\rm lbl}-1.5052806 C_FT_f-3.1454271 C_F^2\nonumber \\
&+&i \left[-5.1387102 C_AC_F+(4.7104510C_F T_f)_{\rm lbl}+5.5811583 C_FT_f+3.7998942C_F^2\right],\nonumber \\
f^{\chi_{c1},(2)}_{0,1}(0,0)&=&-3.1168368 C_AC_F+(4.3673575C_F T_f)_{\rm lbl}-0.5784898C_FT_f-4.4423596 C_F^2\nonumber \\
&+&i \left[-4.5807747C_AC_F+(4.9547864C_F T_f)_{\rm lbl}+6.6283558 C_FT_f+1.0579313C_F^2\right],\nonumber \\
f^{\chi_{c2},(2)}_{2,1}(0,0)&=&-8.9566696 C_AC_F-(4.4272605C_F T_f)_{\rm lbl}+7.9045895 C_FT_f-4.4336479 C_F^2\nonumber \\
&+&i \left[0.8405448 C_AC_F+(2.5927916C_F T_f)_{\rm lbl}-0.6582522 C_FT_f-5.05114415C_F^2\right],\nonumber \\
f^{\chi_{c2},(2)}_{1,1}(0,0)&=&-9.5922329 C_AC_F-(0.3786112C_F T_f)_{\rm lbl}+9.5719362 C_FT_f-6.0608504 C_F^2\nonumber \\
&+&i \left[1.9978170 C_AC_F-(0.8019720C_F T_f)_{\rm lbl}-1.8764477 C_FT_f-6.0703667C_F^2\right],\nonumber \\
f^{\chi_{c2},(2)}_{0,1}(0,0)&=&-6.9298711C_AC_F+(1.8696196C_F T_f)_{\rm lbl}+7.0857442 C_FT_f-2.4006309 C_F^2\nonumber \\
&+&i \left[-2.6301256 C_AC_F+(1.2507569C_F T_f)_{\rm lbl}+4.9939479 C_FT_f-3.0995204C_F^2\right],\nonumber \\
\end{eqnarray}

\begin{eqnarray}\label{one-ex1}
f^{\eta_c,(2)}_{0,1}(1,4)&=&\frac{C_A C_F}{128}-\frac{5C_F^2}{192},\nonumber \\
f^{\chi_{c0},(2)}_{0,1}(1,4)&=&\frac{C_A C_F}{48}-\frac{5C_F^2}{96},\nonumber \\
f^{\chi_{c1},(2)}_{1,1}(1,4)&=&\frac{C_AC_F}{128}-\frac{C_F^2}{192},\nonumber \\
f^{\chi_{c1},(2)}_{0,1}(1,4)&=&\frac{C_A C_F}{128}-\frac{5C_F^2}{192},\nonumber \\
f^{\chi_{c2},(2)}_{2,1}(1,4)&=&\frac{C_A C_F}{48}-\frac{7C_F^2}{192}+\left(\frac{C_FT_f}{32}\right)_{\rm lbl},\nonumber \\
f^{\chi_{c2},(2)}_{1,1}(1,4)&=&\frac{C_A C_F}{128}-\frac{5C_F^2}{192},\nonumber \\
f^{\chi_{c2},(2)}_{0,1}(1,4)&=&\frac{C_A C_F}{48}-\frac{5C_F^2}{96},
\end{eqnarray}

\begin{eqnarray}\label{one-ex1}
f^{\eta_c,(2)}_{0,1}(1,3)&=&\left(\frac{5}{72}-\frac{\ln 2}{32}\right) C_AC_F-\frac{C_FT_f}{18}+\left(-\frac{1}{32}-\frac{\ln 2}{24}\right) C_F^2\nonumber \\
&+&i \left(\frac{1}{32} \pi  C_AC_F-\frac{5 \pi C_F^2}{48}\right),\nonumber \\
f^{\chi_{c0},(2)}_{0,1}(1,3)&=&\left(\frac{1}{3}-\frac{5 \ln 2}{48}\right) C_AC_F-\frac{C_FT_f}{6}+\left(\frac{\ln 2}{6}C_FT_f\right)_{\rm lbl}+\left(-\frac{59}{96}-\frac{\ln 2}{24}\right)C_F^2\nonumber \\
&+&i \left(\frac{1}{12} \pi  C_AC_F-\frac{5 \pi C_F^2}{24}\right),\nonumber \\
f^{\chi_{c1},(2)}_{1,1}(1,3)&=&\left(\frac{\ln 2}{24}-\frac{1}{24}\right) C_AC_F+\left(\frac{17}{96}-\frac{\ln 2}{48}\right)C_F^2\nonumber \\
&+&i \left(\frac{1}{32} \pi  C_AC_F-\frac{\pi C_F^2}{48}\right),\nonumber \\
f^{\chi_{c1},(2)}_{0,1}(1,3)&=&\left(\frac{5}{72}-\frac{\ln 2}{32}\right) C_AC_F-\frac{C_FT_f}{18}+\left(\frac{1}{96}-\frac{\ln 2}{24}\right)C_F^2\nonumber \\
&+&i \left(\frac{1}{32} \pi  C_AC_F-\frac{5 \pi C_F^2}{48}\right),\nonumber \\
f^{\chi_{c2},(2)}_{2,1}(1,3)&=&\frac{\ln 2}{6} C_A C_F-\left(\frac{C_FT_f}{16}\right)_{\rm lbl}+\left(\frac{1}{16}-\frac{\ln 2}{3}\right)C_F^2\nonumber \\
&+&i \left[\frac{\pi}{12}   C_AC_F+\left(\frac{1}{12} \pi C_FT_f\right)_{\rm lbl} -\frac{7\pi}{48}  C_F^2\right],\nonumber \\
f^{\chi_{c2},(2)}_{1,1}(1,3)&=&\left(\frac{1}{4}-\frac{5 \ln 2}{48}\right) C_AC_F-\left(\frac{1}{3} \ln 2 C_F T_f\right)_{\rm lbl}+\left(\frac{13 \ln 2}{48}-\frac{19}{32}\right) C_F^2\nonumber \\
&+&i \left(\frac{1}{32} \pi  C_AC_F-\frac{5 \pi C_F^2}{48}\right),\nonumber \\
f^{\chi_{c2},(2)}_{0,1}(1,3)&=&\left(\frac{1}{3}-\frac{5 \ln 2}{48}\right) C_AC_F-\frac{C_FT_f}{6}+\left(\frac{\ln 2}{6}C_FT_f\right)_{\rm lbl}+\left(-\frac{47}{96}-\frac{\ln 2}{24}\right)C_F^2\nonumber \\
&+&i \left(\frac{1}{12} \pi  C_AC_F-\frac{5 \pi C_F^2}{24}\right),
\end{eqnarray}

\begin{eqnarray}\label{one-ex1}
f^{\eta_c,(2)}_{0,1}(1,2)&=&\left(-\frac{35}{576}-\frac{\pi^2}{12}-\frac{\ln^2 2}{32}-\frac{7 \ln 2}{16}\right) C_AC_F
+\left(\frac{71}{64}+\frac{37 \pi ^2}{192}+\frac{23\ln ^2 2}{32}-\frac{3\ln 2}{32}\right) C_F^2\nonumber\\
&+&\left(\frac{1}{36}+\frac{\ln 2}{2}\right)C_FT_f+\left[\left(\frac{23}{16}-\frac{5 \pi ^2}{16}-\frac{15 \ln ^2 2}{8}+\frac{35 \ln 2}{8}\right)C_FT_f\right]_{\rm lbl} \nonumber\\
&+&i \left[\left(\frac{3 \pi}{32}-\frac{3}{32} \pi \ln 2\right) C_AC_F+\left(-\frac{3 \pi}{32}-\frac{1}{8} \pi  \ln 2\right)C_F^2\right],\nonumber \\
f^{\chi_{c0},(2)}_{0,1}(1,2)&=&\left(-\frac{53}{64}-\frac{7 \pi^2}{64}+\frac{\ln^2 2}{16}-\frac{173 \ln 2}{96}\right) C_AC_F+\left(\frac{11}{64}+\frac{9 \pi ^2}{32}+\frac{23\ln ^2 2}{32}+\frac{119\ln 2}{32}\right) C_F^2\nonumber\\
&+&\left(\frac{5}{4}+\frac{7 \ln 2}{6}\right) C_F T_f+\left[\left(\frac{13}{8}-\frac{17 \pi ^2}{48}-\frac{5 \ln^2 2}{2}+\frac{41\ln 2}{8}\right)C_F
T_f\right]_{\rm lbl} \nonumber\\
&+&i \left[\left(\frac{21 \pi}{32}-\frac{5}{16} \pi \ln 2\right) C_AC_F+\left(\frac{1}{2} \pi  \ln 2 C_F T_f\right)_{\rm lbl} +\left(-\frac{59\pi }{32}-\frac{1}{8} \pi \ln 2\right)C_F^2\right],\nonumber \\
f^{\chi_{c1},(2)}_{1,1}(1,2)&=&\left(-\frac{19}{192}-\frac{71 \pi^2}{384}-\frac{37 \ln^2 2}{64}+\frac{65 \ln 2}{48}\right) C_AC_F+\left(-\frac{51}{64}+\frac{\pi ^2}{3}+\frac{19\ln ^2 2}{32}-\frac{37\ln 2}{32}\right) C_F^2\nonumber\\
&+&\left(\frac{\ln 2}{6}-\frac{1}{12}\right) C_FT_f+\left[\left(-\frac{81}{8}+\frac{5 \pi ^2}{3}+10 \ln^2 2-\frac{63\ln 2}{4}\right)C_F T_f\right]_{\rm lbl} \nonumber\\
&+&i \left[\left(\frac{1}{8} \pi \ln 2-\frac{\pi}{8}\right) C_AC_F+\left(\frac{17 \pi}{32}-\frac{1}{16} \pi \ln 2\right)C_F^2\right],\nonumber \\
f^{\chi_{c1},(2)}_{0,1}(1,2)&=&\left(-\frac{143}{576}-\frac{\pi^2}{24}+\frac{7 \ln^2 2}{32}-\frac{13 \ln 2}{16}\right) C_AC_F+\left(\frac{51}{64}+\frac{7 \pi ^2}{64}+\frac{7\ln ^2 2}{32}+\frac{29\ln 2}{32}\right) C_F^2\nonumber\\
&+&\left(\frac{1}{36}+\frac{\ln 2}{2}\right) C_F T_f+\left[\left(\frac{115}{16}-\frac{55 \pi ^2}{48}-\frac{55 \ln^2 2}{8}+\frac{95 \ln 2}{8}\right)C_FT_f\right]_{\rm lbl} \nonumber\\
&+&i \left[\left(\frac{3 \pi}{32}-\frac{3}{32} \pi \ln 2\right) C_AC_F+\left(\frac{\pi}{32}
-\frac{1}{8} \pi  \ln 2\right)C_F^2\right],\nonumber \\
f^{\chi_{c2},(2)}_{2,1}(1,2)&=&\left(\frac{29}{48}-\frac{7 \pi^2}{64}-\frac{3 \ln^2 2}{16}-\frac{79 \ln 2}{48}\right) C_AC_F+\left(\frac{3 \pi^2}{16}+\frac{69 \ln^2 2}{32}+\frac{15 \ln 2}{32}\right) C_F^2\nonumber\\
&+&\left(\frac{8 \ln 2}{3}-\frac{11}{6}\right) C_F T_f+\left[\left(\frac{153}{16}-\frac{41 \pi^2}{24}-\frac{39 \ln ^2 2}{4}+14 \ln 2\right)C_FT_f\right]_{\rm lbl} \nonumber\\
&+&i \left\{\frac{1}{2} \pi  \ln 2 C_AC_F+\left[\left(\frac{1}{4} \pi \ln 2-\frac{3 \pi}{8}\right)C_FT_f\right]_{\rm lbl} +\left(\frac{3 \pi}{16}-\pi  \ln 2\right)C_F^2\right\},\nonumber \\
f^{\chi_{c2},(2)}_{1,1}(1,2)&=&\left(-\frac{89}{192}-\frac{97 \pi^2}{384}-\frac{19 \ln^2 2}{64}+\frac{41 \ln 2}{12}\right) C_AC_F+\left(\frac{23}{64}+\frac{53 \pi ^2}{96}-\frac{35\ln ^2 2}{32}-\frac{151\ln 2}{32}\right) C_F^2\nonumber\\
&+&\left(\frac{25}{12}-\frac{17 \ln 2}{6}\right) C_F T_f+\left[\left(-\frac{193}{8}+\frac{23 \pi ^2}{6}+\frac{95\ln ^2 2}{4}-\frac{145 \ln 2}{4}\right)C_FT_f\right]_{\rm lbl} \nonumber\\
&+&i \left\{\left(\frac{3 \pi}{4}-\frac{5}{16} \pi  \ln 2\right) C_AC_F+\left[\left(\frac{\pi}{4}-\pi  \ln 2\right)C_F T_f\right]_{\rm lbl}+\left(\frac{13}{16}\pi  \ln 2-\frac{57 \pi}{32}\right)C_F^2\right\},\nonumber \\
f^{\chi_{c2},(2)}_{0,1}(1,2)&=&\left(-\frac{45}{64}+\frac{\pi^2}{64}+\frac{13 \ln^2 2}{16}-\frac{281 \ln 2}{96}\right) C_AC_F+\left(\frac{71}{64}+\frac{\pi ^2}{32}-\frac{25\ln ^2 2}{32}+\frac{179\ln 2}{32}\right) C_F^2\nonumber\\
&+&\left(\frac{1}{4}+\frac{7 \ln 2}{6}\right) C_F T_f+\left[\left(\frac{67}{4}-\frac{137 \pi^2}{48}-\frac{35 \ln ^2 2}{2}+\frac{229 \ln 2}{8}\right) C_FT_f\right]_{\rm lbl}\nonumber\\
&+&i \left\{\left(\frac{21 \pi}{32}-\frac{5}{16} \pi \ln 2\right) C_AC_F+\left[\left(\frac{1}{2} \pi \ln 2-\frac{\pi}{8}\right) C_FT_f\right]_{\rm lbl}+\left(-\frac{47 \pi}{32}-\frac{1}{8} \pi  \ln 2\right)C_F^2\right\},\nonumber\\
\end{eqnarray}

\begin{eqnarray}\label{one-ex1}
f^{\eta_c,(2)}_{0,1}(1,1)&=&\left(\frac{91 \zeta(3)}{128}-\frac{355}{288}-\frac{7 \pi ^2}{576}+\frac{3\ln ^3 2}{16}+\frac{11\ln ^2 2}{48}+\frac{187\ln 2}{144}+\frac{13}{192}\pi ^2 \ln 2\right)C_A C_F\nonumber\\
&+&\left(\frac{179 \zeta(3)}{64}+\frac{37}{32}+\frac{41 \pi ^2}{128}-\frac{35\ln ^3 2}{24}+\frac{15\ln ^2 2}{8}-\frac{107\ln 2}{16}-\frac{29}{96}\pi ^2 \ln 2\right)C_F^2\nonumber\\
&+& \left(\frac{16}{9}-\frac{13 \pi^2}{36}-\frac{\ln^2 2}{3}-\frac{29 \ln 2}{9}\right) C_F T_f\nonumber\\
&+& \left[\left(\frac{165 \zeta(3)}{8}-\frac{103}{8}-\frac{55 \pi ^2}{48}+\frac{5\ln ^3 2}{2}-\frac{37\ln ^2 2}{4}-\frac{23\ln 2}{4}+\frac{5}{8}\pi ^2 \ln 2\right)C_F T_f\right]_{\rm lbl}\nonumber\\
&+&i \left\{\left(-\frac{167 \pi}{288}-\frac{5 \pi^3}{48}-\frac{\pi \ln ^2 2}{16} +\frac{\pi\ln 2}{24} \right) C_AC_F\right.\nonumber\\
&+&\left.\left(\frac{71 \pi}{32}+\frac{17 \pi^3}{96}+\frac{23\pi \ln ^2 2}{16} -\frac{3\pi\ln 2}{16} \right)C_F^2\right.\nonumber\\
&+&\left.\left(\frac{13 \pi}{18}-\frac{1}{3} \pi  \ln 2\right) C_FT_f+\left[\left(\frac{23 \pi}{8}-\frac{5 \pi^3}{8}-\frac{15}{4} \pi \ln ^2 2+\frac{35}{4} \pi\ln 2\right) C_FT_f\right]_{\rm lbl}\right\},\nonumber \\
f^{\chi_{c0},(2)}_{0,1}(1,1)&=&\left(-\frac{\zeta(3)}{8}-\frac{253}{48}-\frac{17 \pi^2}{288}+\frac{\ln^3 2}{8}+\frac{67 \ln^2 2}{48}-\frac{383 \ln 2}{144}+\frac{5}{96} \pi^2 \ln 2\right)C_A C_F\nonumber\\
&+&\left(\frac{93 \zeta(3)}{16}-\frac{89}{32}+\frac{863 \pi ^2}{384}-\frac{25\ln ^3 2}{24}-\frac{91\ln ^2 2}{16}+\frac{143\ln 2}{16}-\frac{37}{96}\pi ^2 \ln 2\right)C_F^2\nonumber\\
&+&\left(\frac{14}{3}-\frac{37 \pi^2}{36}-\frac{2 \ln^2 2}{3}+\frac{31 \ln 2}{9}\right) C_F T_f\nonumber\\
&+&\left[\left(\frac{353 \zeta(3)}{16}-\frac{1}{2}-\frac{79 \pi ^2}{48}+3 \ln^3 2-8 \ln ^2 2-\frac{51\ln 2}{2}+\frac{1}{4}\pi ^2 \ln 2\right)C_F T_f\right]_{\rm lbl}\nonumber\\
&+&i \left\{\left(-\frac{97 \pi}{32}-\frac{5 \pi^3}{96}+\frac{1}{8} \pi \ln ^2 2-\frac{21}{16}\pi  \ln 2\right) C_AC_F\right.\nonumber\\
&+&\left.\left(\frac{11 \pi}{32}+\frac{7 \pi^3}{48}+\frac{23}{16} \pi \ln ^2 2+\frac{119}{16}\pi  \ln 2\right)C_F^2\right.
\nonumber\\
&+&\left.\left(\frac{9 \pi}{2}-\pi  \ln 2\right)C_F T_f+\left[\left(\frac{13 \pi}{4}-\frac{17 \pi ^3}{24}-5\pi  \ln^2 2+\frac{41}{4} \pi \ln 2\right)C_FT_f\right]_{\rm lbl}\right\},\nonumber \\
f^{\chi_{c1},(2)}_{1,1}(1,1)&=&\left(\frac{73 \zeta(3)}{16}-\frac{563}{576}-\frac{61 \pi^2}{144}+\frac{91 \ln^3 2}{96}-\frac{103 \ln^2 2}{24}+\frac{25 \ln 2}{9}+\frac{15}{64} \pi^2 \ln 2\right)C_A C_F\nonumber\\
&+&\left(-\frac{743 \zeta(3)}{64}+\frac{189}{64}+\frac{2 \pi ^2}{3}-\frac{41\ln ^3 2}{48}+\frac{9\ln  ^2 2}{16}-\frac{47\ln 2}{32}-\frac{11}{24}\pi ^2 \ln 2\right)C_F^2\nonumber\\
&+&\left(-\frac{77}{36}-\frac{\pi^2}{36}-\frac{\ln^2 2}{6}+\frac{13 \ln 2}{9}\right)C_F T_f\nonumber\\
&+&\left[\left(-110\zeta(3)+\frac{191}{4}+\frac{149\pi ^2}{24}-\frac{40 \ln^3 2}{3}+53 \ln^2 2+\frac{131 \ln 2}{4}-\frac{10}{3} \pi ^2\ln 2\right)C_F T_f\right]_{\rm lbl}\nonumber\\
&+&i \left\{\left(\frac{3 \pi}{8}-\frac{59 \pi^3}{192}-\frac{37}{32} \pi \ln ^2 2+\frac{47}{16}\pi  \ln 2\right) C_AC_F\right.\nonumber\\
&+&\left.\left(-\frac{51 \pi}{32}+\frac{5 \pi^3}{8}\frac{19}{16} \pi \ln ^2 2-\frac{37}{16}\pi  \ln 2\right)C_F^2\right.\nonumber\\
&+&\left.-\pi C_FT_f+\left[\left(-\frac{81 \pi}{4}+\frac{10 \pi ^3}{3}+20\pi  \ln^2 2-\frac{63}{2} \pi \ln 2\right)C_F T_f\right]_{\rm lbl} \right\},\nonumber 
\end{eqnarray}

\begin{eqnarray}\label{one-ex1}
f^{\chi_{c1},(2)}_{0,1}(1,1)&=&\left(-\frac{261 \zeta(3)}{128}-\frac{49}{96}+\frac{191 \pi ^2}{576}-\frac{7\ln ^3 2}{48}+\frac{53\ln ^2 2}{48}+\frac{301\ln 2}{144}-\frac{1}{64}\pi ^2 \ln 2 \right) C_A C_F \nonumber\\
&+&\left(\frac{531 \zeta(3)}{64}+\frac{21}{32}-\frac{173 \pi ^2}{384}-\frac{19\ln ^3 2}{24}-\frac{5\ln ^2 2}{4}-\frac{127\ln 2}{16}-\frac{13}{96}\pi ^2 \ln 2\right)C_F^2\nonumber\\
&+& \left(\frac{4}{3}-\frac{13 \pi^2}{36}-\frac{\ln^2 2}{3}-\frac{23 \ln 2}{9}\right) C_F T_f\nonumber\\
&+& \left[\left(\frac{605 \zeta(3)}{8}-\frac{217}{8}-\frac{245 \pi ^2}{48}+\frac{55\ln ^3 2}{6}-\frac{167\ln ^2 2}{4}-\frac{43\ln 2}{2}+\frac{55}{24}\pi ^2 \ln 2\right)C_F T_f\right]_{\rm lbl} \nonumber\\
&+&i \left\{\left(-\frac{143 \pi}{288}-\frac{\pi^3}{48}+\frac{7}{16} \pi \ln ^2 2-\frac{17}{24}\pi  \ln 2\right) C_AC_F\right.\nonumber\\
&+&\left.\left(\frac{51 \pi}{32}+\frac{\pi^3}{96}+\frac{7}{16} \pi \ln ^2 2+\frac{29}{16}\pi  \ln 2\right)C_F^2\right.\nonumber\\
&+&\left.\left(\frac{\pi}{18}-\frac{1}{3} \pi  \ln 2\right)+\left[\left(\frac{115 \pi}{8}-\frac{55 \pi^3}{24}-\frac{55}{4} \pi \ln ^2 2+\frac{95}{4} \pi\ln 2\right)C_FT_f\right]_{\rm lbl}\right\},\nonumber\\
f^{\chi_{c2},(2)}_{2,1}(1,1)&=&\left(-\frac{173 \zeta(3)}{32}-\frac{713}{72}+\frac{119 \pi^2}{144}+\frac{\ln^3 2}{16}+\frac{61 \ln^2 2}{48}+\frac{313 \ln 2}{36}-\frac{25}{48} \pi^2 \ln 2\right)C_A C_F\nonumber\\
&+&\left(\frac{47 \zeta(3)}{8}+\frac{949}{64}-\frac{619 \pi ^2}{384}-\frac{59\ln ^3 2}{12}-\frac{\ln^2 2}{2}-\frac{7 \ln 2}{16}+\frac{67}{96} \pi^2 \ln 2\right)C_F^2\nonumber\\
&+&\left(\frac{131}{9}-\frac{4 \pi^2}{9}-\frac{8 \ln^2 2}{3}-\frac{116 \ln 2}{9}\right)C_F T_f\nonumber\\
&+&\left[\left(\frac{1653 \zeta(3)}{16}-\frac{1249}{48}-\frac{173 \pi ^2}{24}+13 \ln^3 2-\frac{119 \ln^2 2}{2}-\frac{649 \ln 2}{24}+\frac{17}{6} \pi^2 \ln 2\right)C_F T_f\right]_{\rm lbl}\nonumber\\
&+&i \left\{\left(-\frac{13 \pi}{12}-\frac{5 \pi^3}{96}-\frac{3}{8} \pi \ln ^2 2+\frac{3}{8} \pi \ln 2\right) C_AC_F\right.\nonumber\\
&+&\left.\left(\frac{\pi^3}{12}+\frac{69}{16} \pi \ln ^2 2+\frac{15}{16}\pi  \ln 2\right)C_F^2\right.\nonumber\\
&+&\left.-\frac{1}{3} \pi C_FT_f+\left[\left(\frac{449 \pi}{24}-\frac{13 \pi^3}{4}-20 \pi  \ln ^2 2+\frac{115}{4} \pi \ln 2\right)C_FT_f\right]_{\rm lbl} \right\},\nonumber \\
f^{\chi_{c2},(2)}_{1,1}(1,1)&=&\left(\frac{635 \zeta(3)}{64}+\frac{6103}{576}-\frac{755 \pi^2}{288}-\frac{\ln^3 2}{32}-\frac{175 \ln^2 2}{24}-\frac{239 \ln 2}{36}+\frac{43}{64} \pi^2 \ln 2\right)C_A C_F\nonumber\\
&+&\left(-\frac{863 \zeta(3)}{64}-\frac{897}{64}+\frac{509 \pi^2}{96}+\frac{205 \ln^3 2}{48}+\frac{249 \ln^2 2}{16}-\frac{369 \ln 2}{32}-\pi ^2 \ln 2\right)C_F^2\nonumber\\
&+&\left(-\frac{431}{36}+\frac{17 \pi^2}{36}+\frac{17 \ln^2 2}{6}+\frac{79 \ln 2}{9}\right) C_F T_f\nonumber\\
&+&\left[\left(-\frac{2003 \zeta(3)}{8}+\frac{305}{6}+\frac{487 \pi ^2}{24}-31 \ln^3 2+159 \ln^2 2+\frac{146 \ln 2}{3}-\frac{27}{4} \pi ^2\ln 2\right)C_F T_f\right]_{\rm lbl} \nonumber\\
&+&i \left\{\left(\frac{41 \pi}{24}-\frac{85 \pi^3}{192}-\frac{19}{32} \pi \ln ^2 2+\frac{47}{16}\pi  \ln 2\right) C_AC_F\right.\nonumber\\
&+&\left.\left(\frac{23 \pi}{32}+\frac{43 \pi^3}{48}-\frac{35}{16} \pi \ln ^2 2-\frac{151}{16}\pi  \ln 2\right)C_F^2\right.\nonumber\\
&+&\left.\frac{1}{3} \pi  C_FT_f+\left[\left(-\frac{571 \pi}{12}+\frac{23 \pi^3}{3}+\frac{95}{2} \pi \ln ^2 2-\frac{147}{2}\pi  \ln 2\right)C_FT_f\right]_{\rm lbl} \right\},\nonumber
\end{eqnarray}

\begin{eqnarray}\label{one-ex1}
f^{\chi_{c2},(2)}_{0,1}(1,1)&=&\left(-\frac{67 \zeta(3)}{8}-\frac{193}{48}+\frac{167 \pi ^2}{144}-\frac{7\ln ^3 2}{8}+\frac{301\ln ^2 2}{48}+\frac{337\ln 2}{144}-\frac{19}{96} \pi^2 \ln 2\right)C_A C_F\nonumber\\
&+&\left(\frac{357 \zeta(3)}{16}+\frac{13}{32}-\frac{73 \pi ^2}{384}+\frac{23\ln ^3 2}{24}-\frac{289\ln ^2 2}{16}-\frac{25\ln 2}{16}+\frac{11}{96}\pi ^2 \ln 2\right)C_F^2\nonumber\\
&+& \left(\frac{2}{3}-\frac{37 \pi^2}{36}-\frac{2 \ln^2 2}{3}-\frac{5 \ln 2}{9}\right)C_F T_f\nonumber\\
&+& \left[\left(\frac{2993 \zeta(3)}{16}-\frac{83}{4}-\frac{823 \pi ^2}{48}+23 \ln^3 2-132 \log^2(2)-\frac{95 \ln 2}{4}+\frac{21}{4} \pi ^2\ln 2\right)C_F T_f\right]_{\rm lbl}\nonumber\\
&+&i \left\{\left(-\frac{89 \pi}{32}+\frac{19 \pi^3}{96}+\frac{13}{8} \pi \ln ^2 2-\frac{57}{16}\pi  \ln 2\right) C_AC_F\right.\nonumber\\
&+&\left.\left(\frac{71 \pi}{32}-\frac{17 \pi^3}{48}-\frac{25}{16} \pi \ln ^2 2+\frac{179}{16}\pi \ln 2\right)C_F^2\right.\nonumber\\
&+&\left.\left(\frac{5 \pi}{2}-\pi  \ln 2\right)C_F T_f+\left[\left(\frac{67 \pi}{2}-\frac{137 \pi^3}{24}-35 \pi  \ln^2 2+\frac{231}{4} \pi \ln 2\right)C_FT_f\right]_{\rm lbl} \right\},\nonumber\\
\end{eqnarray}

\begin{eqnarray}\label{one-ex1}
f^{\eta_c,(2)}_{0,1}(1,0)&=&0.3171111 C_AC_F-(0.5284815C_F T_f)_{\rm lbl}+1.5746514 C_FT_f-6.5406725 C_F^2\nonumber\\
&+&i \left[-0.9057689 C_AC_F-(8.2831888C_F T_f)_{\rm lbl}-6.2063497 C_FT_f+0.6467202C_F^2\right],\nonumber \\
f^{\chi_{c0},(2)}_{0,1}(1,0)&=&17.688766 C_AC_F-(14.235856C_F T_f)_{\rm lbl}-11.702415 C_FT_f-10.483256 C_F^2\nonumber\\
&+&i \left[-9.0952278 C_AC_F-(9.5967000C_F T_f)_{\rm lbl}-15.240290 C_FT_f+44.497487C_F^2\right],\nonumber \\
f^{\chi_{c1},(2)}_{1,1}(1,0)&=&-3.8627561 C_AC_F-(1.0018013C_F T_f)_{\rm lbl}+12.330188 C_FT_f-11.686657 C_F^2\nonumber\\
&+&i \left[3.1589294 C_AC_F-(8.0126701C_F T_f)_{\rm lbl}-0.7995722 C_FT_f-16.900483C_F^2\right],\nonumber \\
f^{\chi_{c1},(2)}_{0,1}(1,0)&=&-4.4803032 C_AC_F-(6.1775727C_F T_f)_{\rm lbl}+6.9615832 C_FT_f-4.9733698 C_F^2\nonumber\\
&+&i \left[2.0996054 C_AC_F-(9.4596090C_F T_f)_{\rm lbl}-5.5082180 C_FT_f-4.6301873C_F^2\right],\nonumber \\
f^{\chi_{c2},(2)}_{2,1}(1,0)&=&-1.2397073 C_AC_F+(16.715151C_F T_f)_{\rm lbl}+3.5300990 C_FT_f-6.2739688 C_F^2\nonumber\\
&+&i \left[-9.1057550 C_AC_F+(6.4225178C_F T_f)_{\rm lbl}-0.0148605 C_FT_f+16.484714C_F^2\right],\nonumber \\
f^{\chi_{c2},(2)}_{1,1}(1,0)&=&-1.0508348 C_AC_F+(2.8951349C_F T_f)_{\rm lbl}-2.6657509 C_FT_f+3.6256517 C_F^2\nonumber\\
&+&i \left[-10.299476 C_AC_F+(3.6608930C_F T_f)_{\rm lbl}+0.0445385 C_FT_f+25.621853C_F^2\right],\nonumber \\
f^{\chi_{c2},(2)}_{0,1}(1,0)&=&2.7990579 C_AC_F-(5.6812085C_F T_f)_{\rm lbl}-6.6897024 C_FT_f-11.309981 C_F^2\nonumber\\
&+&i \left[-0.1576692 C_AC_F-(4.3048316C_F T_f)_{\rm lbl}-16.953015 C_FT_f+20.290327C_F^2\right].\nonumber\\
\end{eqnarray}




\begin{thebibliography}{150}

\bibitem{Abe:2002rb}
K.~Abe \textit{et al.} [Belle],
Phys. Rev. Lett. \textbf{89}, 142001 (2002)
doi:10.1103/PhysRevLett.89.142001
[arXiv:hep-ex/0205104 [hep-ex]].

\bibitem{Aubert:2005tj}
B.~Aubert \textit{et al.} [BaBar],
Phys. Rev. D \textbf{72}, 031101 (2005)
doi:10.1103/PhysRevD.72.031101
[arXiv:hep-ex/0506062 [hep-ex]].

\bibitem{Pakhlov:2009nj}
P.~Pakhlov \textit{et al.} [Belle],
Phys. Rev. D \textbf{79}, 071101 (2009)
doi:10.1103/PhysRevD.79.071101
[arXiv:0901.2775 [hep-ex]].

\bibitem{Belle:2018jqa}
S.~Jia \textit{et al.} [Belle],
Phys. Rev. D \textbf{98}, no.9, 092015 (2018)
doi:10.1103/PhysRevD.98.092015
[arXiv:1810.10291 [hep-ex]].

\bibitem{BESIII:2014uzr}
M.~Ablikim \textit{et al.} [BESIII],
Chin. Phys. C \textbf{39}, no.4, 041001 (2015)
doi:10.1088/1674-1137/39/4/041001
[arXiv:1411.6336 [hep-ex]].

\bibitem{BESIII:2017nty}
M.~Ablikim \textit{et al.} [BESIII],
Phys. Rev. D \textbf{96}, no.5, 051101 (2017)
doi:10.1103/PhysRevD.96.051101
[arXiv:1705.06853 [hep-ex]].

\bibitem{Bodwin:1994jh}
G.~T.~Bodwin, E.~Braaten and G.~P.~Lepage,
Phys. Rev. D \textbf{51}, 1125-1171 (1995)
[erratum: Phys. Rev. D \textbf{55}, 5853 (1997)]
doi:10.1103/PhysRevD.55.5853
[arXiv:hep-ph/9407339 [hep-ph]].

\bibitem{Chung:2008km}
H.~S.~Chung, J.~Lee and C.~Yu,
Phys. Rev. D \textbf{78}, 074022 (2008)
doi:10.1103/PhysRevD.78.074022
[arXiv:0808.1625 [hep-ph]].

\bibitem{Sang:2009jc}
W.~L.~Sang and Y.~Q.~Chen,
Phys. Rev. D \textbf{81}, 034028 (2010)
doi:10.1103/PhysRevD.81.034028
[arXiv:0910.4071 [hep-ph]].

\bibitem{Li:2009ki}
D.~Li, Z.~G.~He and K.~T.~Chao,
Phys. Rev. D \textbf{80}, 114014 (2009)
doi:10.1103/PhysRevD.80.114014
[arXiv:0910.4155 [hep-ph]].

\bibitem{Wang:2013ywc}
X.~P.~Wang and D.~Yang,
JHEP \textbf{06}, 121 (2014)
doi:10.1007/JHEP06(2014)121
[arXiv:1401.0122 [hep-ph]].

\bibitem{Chung:2019ota}
H.~S.~Chung, J.~H.~Ee, D.~Kang, U.~R.~Kim, J.~Lee and X.~P.~Wang,
JHEP \textbf{10}, 162 (2019)
doi:10.1007/JHEP10(2019)162
[arXiv:1906.03275 [hep-ph]].

\bibitem{Fan:2012dy}
Y.~Fan, J.~Lee and C.~Yu,
Phys. Rev. D \textbf{87}, no.9, 094032 (2013)
doi:10.1103/PhysRevD.87.094032
[arXiv:1211.4111 [hep-ph]].

\bibitem{Xu:2014zra}
G.~Z.~Xu, Y.~J.~Li, K.~Y.~Liu and Y.~J.~Zhang,
JHEP \textbf{10}, 071 (2014)
doi:10.1007/JHEP10(2014)071
[arXiv:1407.3783 [hep-ph]].

\bibitem{Brambilla:2017kgw}
N.~Brambilla, W.~Chen, Y.~Jia, V.~Shtabovenko and A.~Vairo,
Phys. Rev. D \textbf{97}, no.9, 096001 (2018)
[erratum: Phys. Rev. D \textbf{101}, no.3, 039903 (2020)]
doi:10.1103/PhysRevD.97.096001
[arXiv:1712.06165 [hep-ph]].

\bibitem{Chen:2017pyi}
L.~B.~Chen, Y.~Liang and C.~F.~Qiao,
JHEP \textbf{01}, 091 (2018)
doi:10.1007/JHEP01(2018)091
[arXiv:1710.07865 [hep-ph]].

\bibitem{Yu:2020tri}
H.~M.~Yu, W.~L.~Sang, X.~D.~Huang, J.~Zeng, X.~G.~Wu and S.~J.~Brodsky,
JHEP \textbf{01}, 131 (2021)
doi:10.1007/JHEP01(2021)131
[arXiv:2007.14553 [hep-ph]].

\bibitem{Sang:2020fql}
W.~L.~Sang, F.~Feng and Y.~Jia,
JHEP \textbf{10}, 098 (2020)
doi:10.1007/JHEP10(2020)098
[arXiv:2008.04898 [hep-ph]].

\bibitem{Jia:2008ep}
Y.~Jia and D.~Yang,
Nucl. Phys. B \textbf{814}, 217-230 (2009)
doi:10.1016/j.nuclphysb.2009.01.025
[arXiv:0812.1965 [hep-ph]].

\bibitem{Braguta:2010mf}
V.~V.~Braguta,
Phys. Rev. D \textbf{82}, 074009 (2010)
doi:10.1103/PhysRevD.82.074009
[arXiv:1006.5798 [hep-ph]].

\bibitem{Trunin:2024ibx}
A.~Trunin,
[arXiv:2406.05729 [hep-ph]].

\bibitem{Liao:2021ifc}
Q.~L.~Liao, J.~Jiang, P.~C.~Lu and G.~Chen,
Phys. Rev. D \textbf{105}, no.1, 016026 (2022)
doi:10.1103/PhysRevD.105.016026
[arXiv:2112.03522 [hep-ph]].

\bibitem{Chen:2013itc}
G.~Chen, X.~G.~Wu, Z.~Sun, X.~C.~Zheng and J.~M.~Shen,
Phys. Rev. D \textbf{89}, no.1, 014006 (2014)
doi:10.1103/PhysRevD.89.014006
[arXiv:1311.2735 [hep-ph]].

\bibitem{Chen:2013mjb}
G.~Chen, X.~G.~Wu, Z.~Sun, S.~Q.~Wang and J.~M.~Shen,
Phys. Rev. D \textbf{88}, 074021 (2013)
doi:10.1103/PhysRevD.88.074021
[arXiv:1308.5375 [hep-ph]].

\bibitem{Chao:2013cca}
K.~T.~Chao, Z.~G.~He, D.~Li and C.~Meng,
[arXiv:1310.8597 [hep-ph]].

\bibitem{Li:2013nna}
Y.~J.~Li, G.~Z.~Xu, K.~Y.~Liu and Y.~J.~Zhang,
JHEP \textbf{01}, 022 (2014)
doi:10.1007/JHEP01(2014)022
[arXiv:1310.0374 [hep-ph]].

\bibitem{Feng:2015uha}
F.~Feng, Y.~Jia and W.~L.~Sang,
Phys. Rev. Lett. \textbf{115}, no.22, 222001 (2015)
doi:10.1103/PhysRevLett.115.222001
[arXiv:1505.02665 [hep-ph]].

\bibitem{Wang:2018lry}
S.~Q.~Wang, X.~G.~Wu, W.~L.~Sang and S.~J.~Brodsky,
Phys. Rev. D \textbf{97}, no.9, 094034 (2018)
doi:10.1103/PhysRevD.97.094034
[arXiv:1804.06106 [hep-ph]].

\bibitem{Babiarz:2025agk}
I.~Babiarz, C.~A.~Flett, M.~A.~Ozcelik, W.~Sch{\"a}fer and A.~Szczurek,
[arXiv:2509.15310 [hep-ph]].

\bibitem{Bodwin:2002cfe}
G.~T.~Bodwin and A.~Petrelli,
Phys. Rev. D \textbf{66}, 094011 (2002)
[erratum: Phys. Rev. D \textbf{87}, no.3, 039902 (2013)]
doi:10.1103/PhysRevD.66.094011
[arXiv:hep-ph/0205210 [hep-ph]].

\bibitem{Zhang:2021ted}
Y.~D.~Zhang, F.~Feng, W.~L.~Sang and H.~F.~Zhang,
JHEP \textbf{12}, 189 (2021)
doi:10.1007/JHEP12(2021)189
[arXiv:2109.15223 [hep-ph]].

\bibitem{Czarnecki:1997vz}
A.~Czarnecki and K.~Melnikov,
Phys. Rev. Lett. \textbf{80}, 2531-2534 (1998)
doi:10.1103/PhysRevLett.80.2531
[arXiv:hep-ph/9712222 [hep-ph]].

\bibitem{Beneke:1997jm}
M.~Beneke, A.~Signer and V.~A.~Smirnov,
Phys. Rev. Lett. \textbf{80}, 2535-2538 (1998)
doi:10.1103/PhysRevLett.80.2535
[arXiv:hep-ph/9712302 [hep-ph]].

\bibitem{Sang:2015uxg}
W.~L.~Sang, F.~Feng, Y.~Jia and S.~R.~Liang,
Phys. Rev. D \textbf{94}, no.11, 111501 (2016)
doi:10.1103/PhysRevD.94.111501
[arXiv:1511.06288 [hep-ph]].

\bibitem{Hoang:2006ty}
A.~H.~Hoang and P.~Ruiz-Femenia,
Phys. Rev. D \textbf{74}, 114016 (2006)
doi:10.1103/PhysRevD.74.114016
[arXiv:hep-ph/0609151 [hep-ph]].

\bibitem{Hahn:2000kx}
T.~Hahn,
Comput. Phys. Commun. \textbf{140}, 418-431 (2001)
doi:10.1016/S0010-4655(01)00290-9
[arXiv:hep-ph/0012260 [hep-ph]].

\bibitem{Mertig:1990an}
R.~Mertig, M.~Bohm and A.~Denner,
Comput. Phys. Commun. \textbf{64}, 345-359 (1991)
doi:10.1016/0010-4655(91)90130-D

\bibitem{Feng:2012tk}
F.~Feng and R.~Mertig,
[arXiv:1212.3522 [hep-ph]].

\bibitem{Beneke:1997zp}
M.~Beneke and V.~A.~Smirnov,
Nucl. Phys. B \textbf{522}, 321-344 (1998)
doi:10.1016/S0550-3213(98)00138-2
[arXiv:hep-ph/9711391 [hep-ph]].

\bibitem{CalcLoop}
The package {\tt CalcLoop}: https://gitlab.com/multiloop-pku/calcloop.

\bibitem{Klappert:2020nbg}
J.~Klappert, F.~Lange, P.~Maierh{\"o}fer and J.~Usovitsch,
Comput. Phys. Commun. \textbf{266}, 108024 (2021)
doi:10.1016/j.cpc.2021.108024
[arXiv:2008.06494 [hep-ph]].

\bibitem{Guan:2024byi}
X.~Guan, X.~Liu, Y.~Q.~Ma and W.~H.~Wu,
Comput. Phys. Commun. \textbf{310}, 109538 (2025)
doi:10.1016/j.cpc.2025.109538
[arXiv:2405.14621 [hep-ph]].

\bibitem{Smirnov:2014hma}
A.~V.~Smirnov,
Comput. Phys. Commun. \textbf{189}, 182-191 (2015)
doi:10.1016/j.cpc.2014.11.024
[arXiv:1408.2372 [hep-ph]].

\bibitem{Binosi:2008ig}
D.~Binosi, J.~Collins, C.~Kaufhold and L.~Theussl,
Comput. Phys. Commun. \textbf{180}, 1709-1715 (2009)
doi:10.1016/j.cpc.2009.02.020
[arXiv:0811.4113 [hep-ph]].

\bibitem{Liu:2017jxz}
X.~Liu, Y.~Q.~Ma and C.~Y.~Wang,
Phys. Lett. B \textbf{779}, 353-357 (2018)
doi:10.1016/j.physletb.2018.02.026
[arXiv:1711.09572 [hep-ph]].

\bibitem{Liu:2022mfb}
Z.~F.~Liu and Y.~Q.~Ma,
Phys. Rev. Lett. \textbf{129}, no.22, 222001 (2022)
doi:10.1103/PhysRevLett.129.222001
[arXiv:2201.11637 [hep-ph]].

\bibitem{Liu:2022chg}
X.~Liu and Y.~Q.~Ma,
Comput. Phys. Commun. \textbf{283}, 108565 (2023)
doi:10.1016/j.cpc.2022.108565
[arXiv:2201.11669 [hep-ph]].

\bibitem{Li:2025mng}
C.~Li, X.~D.~Huang and W.~L.~Sang,
[arXiv:2506.16317 [hep-ph]].

\bibitem{Chen:2025qgy}
X.~Chen, X.~Guan, C.~Q.~He, Y.~Q.~Ma, J.~Wang and D.~J.~Zhang,
[arXiv:2508.20777 [hep-ph]].

\bibitem{Chen:2025ocd}
X.~Chen, X.~Guan, C.~Q.~He and Y.~Q.~Ma,
[arXiv:2510.10261 [hep-ph]].

\bibitem{Sang:2023hjl}
W.~L.~Sang, D.~S.~Yang and Y.~D.~Zhang,
Phys. Rev. D \textbf{108}, no.1, 014021 (2023)
doi:10.1103/PhysRevD.108.014021
[arXiv:2302.06439 [hep-ph]].

\bibitem{Sang:2022erv}
W.~L.~Sang, D.~S.~Yang and Y.~D.~Zhang,
Phys. Rev. D \textbf{106}, no.9, 094023 (2022)
doi:10.1103/PhysRevD.106.094023
[arXiv:2208.10118 [hep-ph]].

\bibitem{Chetyrkin:2000yt}
K.~G.~Chetyrkin, J.~H.~Kuhn and M.~Steinhauser,
Comput. Phys. Commun. \textbf{133}, 43-65 (2000)
doi:10.1016/S0010-4655(00)00155-7
[arXiv:hep-ph/0004189 [hep-ph]].

\bibitem{Bodwin:2007fz}
G.~T.~Bodwin, H.~S.~Chung, D.~Kang, J.~Lee and C.~Yu,
Phys. Rev. D \textbf{77}, 094017 (2008)
doi:10.1103/PhysRevD.77.094017
[arXiv:0710.0994 [hep-ph]].

\bibitem{Bodwin:2006dn}
G.~T.~Bodwin, D.~Kang and J.~Lee,
Phys. Rev. D \textbf{74}, 014014 (2006)
doi:10.1103/PhysRevD.74.014014
[arXiv:hep-ph/0603186 [hep-ph]].

\end{thebibliography}

\end{document}